\documentclass[12pt]{iopart}
\usepackage{graphicx,psfrag}
\usepackage{color}
\usepackage{iopams}
\usepackage{amssymb}

\newcommand{\ra}{\rangle}
\newcommand{\la}{\langle}

\newcommand{\mc}{\mathcal}

\newcommand{\be}{\begin{equation}}
\newcommand{\ee}{\end{equation}}
\newcommand{\bea}{\begin{eqnarray}}
\newcommand{\eea}{\end{eqnarray}}
\newcommand{\ba}{\begin{array}}
\newcommand{\ea}{\end{array}}
\newcommand{\ben}{\begin{enumerate}}
\newcommand{\een}{\end{enumerate}}

\newcommand{\bt}{\begin{table}}
\newcommand{\et}{\end{table}}
\newcommand{\btr}{\begin{tabular}}
\newcommand{\etr}{\end{tabular}}
\newcommand{\bfi}{\begin{figure}}
\newcommand{\efi}{\end{figure}}

\begin{document}

\title{Completeness of classical spin models and universal quantum computation}
\author{Gemma De las Cuevas$^{\dag\ddag}$,  Wolfgang D\"ur$^{\dag \ddag}$, Maarten Van den Nest$^{\S}$ and Hans J. Briegel$^{\dag\ddag}$ }
\address{$^\dag$Institut f\"ur Quantenoptik und Quanteninformation der \"Osterreichischen Akademie der Wissenschaften, Innsbruck, Austria}
\address{$^\ddag$Institut f{\"u}r Theoretische Physik, Universit{\"a}t
Innsbruck, Technikerstra{\ss}e 25, A-6020 Innsbruck,
Austria}
\address{$^\S$Max-Planck Institut f\"ur Quantenoptik, Hans-Kopfermann-Str.~1, D-85748 Garching, Germany}

\begin{abstract}
We study mappings between distinct classical spin systems that leave the partition function invariant. As recently shown in [\emph{Phys. Rev. Lett.} {\bf 100} 110501 (2008)], the partition function of the 2D square lattice Ising model in the presence of an inhomogeneous magnetic field, can specialize to the partition function of any Ising system on an arbitrary graph. In this sense the 2D Ising model is said to be ``complete''. However, in order to obtain the above result, the coupling strengths on the 2D lattice must assume complex values, and thus do not allow for a physical interpretation. Here we show how a complete model with real ---and, hence, ``physical''--- couplings can be obtained if the 3D Ising model is considered. We furthermore show how to map general $q$-state systems with possibly many-body interactions to the 2D Ising model with complex parameters, and give completeness results for these models with real parameters. We also demonstrate that the computational overhead in these constructions is in all relevant cases polynomial. These results are proved by invoking a recently found cross-connection between statistical mechanics and  quantum information theory, where partition functions are expressed as quantum mechanical amplitudes. Within this framework, there exists a natural correspondence between many-body quantum states that allow universal quantum computation via local measurements only, and complete classical spin systems.
\end{abstract}

\maketitle

\tableofcontents

\section{Introduction}

Classical spin systems, such as the Ising and Potts models~\cite{Wu84}, are widely studied in physics and appear in different contexts, ranging from magnetism to economic models~\cite{Me87}. Even in two spatial dimensions, these systems can show interesting and complex features ---as is apparent in e.g.~their phase diagrams--- and they can be difficult to treat analytically or numerically~\cite{Ba82}.

One of the central characteristics of a spin system is its partition function, as many relevant macroscopic properties of the system can be derived from it. In this paper we concentrate on mappings between classical spin systems of possibly different size that leave the partition function invariant, and thus also the thermodynamical properties of the system.  On the one hand, we will identify different models that have the same partition function and are hence in this sense equivalent. On the other hand, we will find that there exist certain models that are {\em complete}. Colloquially speaking, a spin model ${\cal M}$, defined on a certain class of lattices and exhibiting a certain range of coupling strengths, is said to be ``complete'' if  the partition function of every other classical spin system ${\cal M}'$ can be re-expressed as a partition function of the model ${\cal M}$, where the lattice size of ${\cal M}$ as well as its couplings depend on ${\cal M}'$~\footnote{In this context we sometimes say that the partition function of the complete model ``specializes'' to the partition function of all other models.}. Such completeness results were first established in~\cite{Va08}, where it was shown that the 2D Ising model with inhomogeneous magnetic fields constitutes a complete model with respect to all $q-$state Potts-type models with pairwise interactions. That is, the partition function of every such Potts-type system (with arbitrary finite $q$, defined on an arbitrary graph and with arbitrary couplings) can be re-expressed as the partition function of an Ising model on a 2D square lattice which is polynomially larger than the original system, with suitably tuned parameters~\footnote{If the completeness only holds for a restricted set of models on certain graphs, it is specified by saying, e.g., ``completeness for Ising models''.}.

The main tool to obtain these results (as first used in~\cite{Va08}) is a new connection between classical statistical mechanics and quantum information theory (see also~\cite{Br07,Bo08,Va07,So07,Li04,Li97,Mu05,Ve06,Ge08,Ah07,Ah06,Ah06a,Wo06,Ar08}, where similar connections are studied). In~\cite{Va08} the partition function of a classical spin system is expressed as a quantum mechanical object, namely a quantum amplitude.  More specifically, partition functions $Z$ are shown to coincide with certain overlaps (inner products) between pairs of many-body quantum states: $Z = \langle\alpha|\psi\rangle$. Here $|\alpha\rangle$ is a ``product state'', i.e.~a state which does not contain any entanglement, whereas $|\psi\rangle$ is a  (typically highly entangled) ``quantum stabilizer state''. Quantum overlaps of the above form appear naturally in a branch of quantum information theory called { ``measurement-based quantum computation''}~\cite{Ra01} (or ``one-way quantum computation''). This a paradigm for quantum computation in which  computations are carried out by performing sequences of (adaptive) single-qubit measurements on a highly entangled many-body ``resource state''. It is known that there exist certain stabilizer states, such as the two--dimensional (2D) cluster states~\cite{Br01} that are {\em universal} states for measurement-based quantum computation. That is, every quantum computation can be carried out by merely performing a suitable sequence of single-qubit measurements on such universal states.  It is this universality feature of the 2D cluster state, together with the aforementioned connection between stabilizer states and classical spin systems, that is used in \cite{Va08} to establish the completeness of the 2D Ising model.

In the present paper, this connection to quantum information theory will be further studied and exploited to strengthen the results of \cite{Va08}. The completeness results presented in \cite{Va08} exhibited some limitations, which will be here significantly relaxed. For example, in \cite{Va08} the completeness of the 2D Ising model could only be obtained when allowing for negative-- or complex--valued ---instead of real and positive--- Boltzmann weights, making it hence difficult to give a physical interpretation of this result. Furthermore, efficient reductions ---i.e.~requiring only a polynomial amount of operations in the system size--- were in \cite{Va08} only provided from Potts-type models to the 2D Ising model; for models which were not of Potts-type (for example, vertex models such as the six-vertex model, which involve four-body interactions), it was unclear whether such reductions to the 2D Ising model could be made efficient as well.  In the present paper we show that  the above limitations can be overcome in several cases. The following is a summary of our main results.
\begin{itemize}
\item
The  Ising model with inhomogeneous magnetic fields on a 2D square lattice (2D Ising model) is complete with complex parameters for {\em all} spin models.  This includes (edge and vertex) models on arbitrary graphs (hence including lattices of arbitrary dimension) with $q-$level spins and $k-$body interactions, for every finite $q$ and $k$. The overhead (i.e.~the size of the 2D Ising model as compared to the original system) is polynomial in the number of spins, as long as $q$ and $k$ are bounded (i.e.~arbitrary, but fixed).
\item
The 3D Ising model without magnetic fields is complete with real parameters for all Ising models with magnetic fields on arbitrary graphs, and only a polynomial overhead is required.
Similarly, we present a $q-$state edge model on a 3D square lattice which is complete with real parameters for all $q'-$state edge models with pairwise interactions as long as $q'\leq q$.
\item Going beyond Ising models as complete models, we extend the analysis to vertex models and we construct a certain 2D two--state vertex model which is complete for all models when allowing for complex parameters. When restricting to the real parameter regime, we construct a 2D $q-$state vertex model which is complete for all $q'-$state vertex and edge models with at most $4-$body interactions with $q'\leq q$. Again, only a polynomial overhead is required.

\item We emphasize that the completeness results  presented here are based on {\em explicit constructions} (which are, as mentioned, in most cases also computationally \emph{efficient}), which we illustrate with help of a number of examples. For instance, we show explicitly how to express the partition function of a 3D Ising model as the partition function of a polynomially enlarged 2D Ising model with magnetic fields.
\end{itemize}

This paper is organized as follows. First we briefly review the quantum formulations of the partition function presented in~\cite{Hu08} (section~\ref{sec:Zinaquantumlanguage}). Then we present completeness results with complex parameters (section~\ref{sec:completeness-2DIsing}); that is, we show that the 2D Ising model is complete with complex parameters for all models, and that a 2D two--state vertex model is also complete with complex parameters for all models. In section~\ref{sec:completeness-real} we present completeness results with real parameters: first we show that the 3D Ising model without magnetic fields is complete with real parameters for Ising models on arbitrary graphs, then we show that a 3D $q-$state edge model is complete with real parameters for $q'-$state edge models with pairwise interactions, with $q'\leq q$, and finally we prove that a 2D $q-$state vertex model is complete with real parameters for all $q'-$state edge and vertex models with at most $4-$body interactions, with $q'\leq q$. In section~\ref{sec:complexity} we give some implications of our results in complexity theory, and finally we conclude in section~\ref{sec:conclusions}.

\section{Quantum formulation of the partition function}
\label{sec:Zinaquantumlanguage}

Here we briefly review four mappings thoroughly exposed in \cite{Hu08} that allow us to express the partition function of a system of classical spins in a quantum language; more precisely, as the overlap between a stabilizer state and a complete product state. We present mappings for the partition function of the Ising model with magnetic fields (introduced in \cite{Va08}), a general class of $q-$state edge models (introduced in \cite{Va07}), both of them in the \emph{Graph picture}, edge models in the \emph{GHZ picture} and vertex models in the \emph{PEPS picture} (both introduced in \cite{Hu08}).

\subsection{Edge models with magnetic fields in the Graph picture}
\label{ssec:phi}

Generally, an edge model represented on a graph is a model in which particles are sitting at the vertices of a graph $G=(V,E)$ and edges represent interactions between these particles, thereby resulting in the \emph{Graph picture} of the edge model.

Here we consider a classical spin model involving $N$ $q-$state spins
$(s_1,s_2\ldots s_N)\equiv {\bf s}$, where $s_a = 0, 1, \ldots, q-1$ and $a=1,\ldots,N$. The spins interact pairwise according to an interaction pattern specified by a
graph $G=(V,E)$ with inhomogeneous coupling strengths $J_{ab}$ and local magnetic fields $h_a$ as
\be
H_G({\bf s})=-\sum _{\{a, b\} \in E} J_{ab}
(2\delta(s_{a},s_{b}) -1) - \sum_{a \in V} h_a s_a,
\label{eq:H-phi}
\ee
where $\delta(s_{a},s_{b})=1$ if $s_a=s_b$ and it is 0 otherwise. This corresponds to the Hamilton function of the Potts model with magnetic fields, which specializes to the Ising model with magnetic fields for $q=2$, as considered in~\cite{Va08}. In the remainder of this subsection we shall focus on the case $q=2$.
The partition function $Z_G$ is defined as
\begin{equation}
 Z_{G}(\{J_{ab},h_a\}) =\sum_{\mathbf{s}} e^{-\beta H_G({\mathbf{s}})},
\end{equation}
where $\beta=1/(k_BT)$,
with $k_B$ the Boltzmann constant and $T$ the temperature.

\begin{figure}[htb]
\centering
\psfrag{(a)}{(a)}
\psfrag{(b)}{(b)}
\psfrag{(c)}{(c)}
\psfrag{a}{$G$}
\psfrag{b}{$|\varphi_G\ra$}
\psfrag{c}{$|\psi_G\ra$}
\includegraphics[width=0.6\columnwidth]{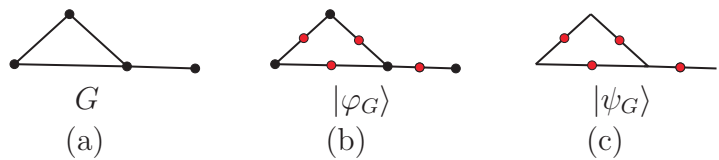}
\caption{
(a) A graph $G=(V,E)$ is used to indicate that classical spins sit at the vertices of this graph and interact with other vertices to which they are connected by an edge.
(b) The quantum state $|\varphi_{G}\ra$ is obtained by placing quantum spins at the vertices of this graph, thereby characterizing the local magnetic fields acting on the corresponding classical spin, and quantum spins on the edges, thereby characterizing the interaction between classical spins along that edge.
(c) The quantum state $|\psi_G\ra$ associated with the graph $G$ places quantum spins only at the edges of $G$, used to characterize the interactions of the classical spins.
}
\label{fig:graph-decoratedgraph}
\end{figure}

To express this partition function in a quantum language, we first associate with each graph  $G$ its decorated version by adding a new vertex at the middle of every edge (see figure~\ref{fig:graph-decoratedgraph}(b)). We call this new set of vertices ``edge qubits'' in contrast to the original set of vertices, called ``vertex qubits''. An edge qubit placed at edge $(a,b)\in E$ is labelled $ab$, thus indicating the two vertex qubits $a$ and $b$ which are its neighbors.
We now define a \emph{stabilizer state} $|\varphi_G\ra$ on the decorated graph:
\begin{equation}
|\varphi_G\ra=\sum_{\bf s}\bigotimes_{ab}|s_a + s_b\ra \bigotimes_{a}|s_a\ra
\label{eq:varphi}
\end{equation}
(for notational transparency we omit the particle outside the ket, that is, we write $|s_a\ra$ instead of $|s_a\ra_a$).
A stabilizer state is the common unique eigenstate of a set of (Pauli) operators; in particular, the state $|\varphi_G\ra$ is left invariant by the operator $\sigma_x^{(a)}\bigotimes_{b:(a,b)\in E}\sigma_x^{(b)}$ for all $a\in V$ and by $\sigma_z^{(ab)}\sigma_z^{(a)}\sigma_z^{(b)}$ for all $(a,b)\in E$, where $\sigma_x$ and $\sigma_z$ are the Pauli matrices (see \cite{Hu08} for the details). Thus, by applying Hadamard rotations to all edge qubits one obtains a graph state (as defined in, e.g., \cite{He06,He04}) on the decorated graph.
We see that this stabilizer state assigns to the vertex qubits the value of the classical spin, and to edge qubits the sum of the values of the two neighboring vertex qubits, and sums over all spin configurations.
In other words, classical spins are associated with the vertices of a graph, and we place quantum degrees of freedom at the vertices of the graph (vertex qubits) to characterize the degrees of freedom of the local magnetic fields of the corresponding classical spins, and quantum degrees of freedom at the edges (edge qubits) to characterize the interactions between classical spins.

We further define a complete product state $|\alpha\ra=\bigotimes_{(a,b)\in E}|\alpha_{ab}\ra \bigotimes_{a\in V}|\alpha_a\ra$, in which
\begin{eqnarray}\label{eq:alpha-phi-a}
|\alpha_a\ra &=& e^{\beta h_a}|0\ra + e^{-\beta h_a}|1\ra
\end{eqnarray}
refers to vertex qubit $a$ and contains the information of the magnetic field at that site, and
\begin{eqnarray}
\label{eq:alpha-phi-ab}
|\alpha_{ab}\ra &=& e^{\beta J_{ab}}|0\ra + e^{-\beta J_{ab}}|1\ra
\end{eqnarray}
refers to the edge qubit $ab$ and contains the information of the coupling strength between spins at vertices $a$ and $b$. Here, $|0\ra$ and $|1\ra$ are defined as the eigenstates of $\sigma_z$ with eigenvalues $+1$ and $-1$, respectively.

The partition function of this system of classical spins can be expressed in a quantum language by computing the overlap
\be
Z_{G}= \langle \alpha|\varphi_{G}\rangle .
\label{eq:Z=alpha|phi}
\ee

\subsection{Edge models without magnetic fields in the Graph picture}
\label{ssec:psi}

In~\cite{Va07}, and in more detail in~\cite{Hu08}, a similar class of spin models is considered, but there the $q-$state particles are not subjected to magnetic fields, and they interact pairwise according to a more general Hamilton function. More precisely, an oriented version of a graph $G$ is considered, $G^{\sigma}$, where, for
every edge $e$, one end-vertex $v_e^+$ is assigned to be the
\emph{head} of $e$, and the other end-vertex $v_e^-$ is the
\emph{tail} of $e$.
The Hamilton function of the system is then given by
\begin{equation}
H_G({\bf s})=\sum _{e \in E} h_e(|s_{v_e^+} - s_{v_e^-}|_q).
\end{equation}
Here every $h_e$ is some local Hamilton function
defined on the edge $e$ with the only restriction that the
interaction strength between the spins on the endpoints
$v_e^+$ and $v_e^-$ is a function of the difference
between $s_{v_e^+}$ and $s_{v_e^-}$ modulo $q$, denoted by
$|s_{v_e^+} - s_{v_e^-}|_q$.
Note that the pairwise interactions here are more general than in~\eref{eq:H-phi}, and they can also specialize to the Potts
model or the Ising model.

We define a stabilizer state $|\psi\ra = \sum_{s\in \mathbb{Z}_q^n}\bigotimes_{ab} ||s_a-s_b|_q\ra$. In order to write its stabilizers, we need the definition of the shift and phase operators, $X$ and $Z$, respectively:
\be
\label{eq:XZ}
X|j\rangle = |j+1\ \mbox{mod }q\rangle,\quad
\mbox{ and }\quad Z|j\rangle = e^{2\pi ij/q}|j\rangle,
\ee
for every $j=0, \dots, q-1$. Then the stabilizer group of $|\psi_G\ra$ is generated by $\bigotimes_{b:(a,b)\in E}X_{ab}^{\tau_{ab}}$ for all $a\in V$, and $\bigotimes_{(a,b)\in \partial f}Z_{ab}^{\nu_{ab}}$ for all faces $f$ (by face we mean a plaquette with the minimal number of edges in its boundary). Here $\tau_{ab}=1 (-1)$ if $a$ is the head (tail) of edge $ab$,  $(a,b) \in\partial f$ means that edge $ab$ is at the boundary of face $f$, and $\nu_{ab} = 1(-1)$ if the oriented edge points clockwise (counterclockwise) according to the center of the face (see \cite{Hu08} for a more detailed analysis).
We see that the state $|\psi_G\ra$ is obtained by placing $q-$dimensional quantum spins at the edges of the  graph $G$, which contain the degrees of freedom of the interaction between the two classical spins at the end points of the edge (see figure~\ref{fig:graph-decoratedgraph}(c)).

We also define a complete product state  $|\alpha \ra= \bigotimes_{e\in E}|\alpha_e\rangle,$ where
\begin{equation}
\label{eq:alpha-psi}
|\alpha_e\rangle = \sum_{j=0}^{q-1} e^{-\beta h_e(j)}|j\rangle
\end{equation}
 is a $q-$dimensional state associated with the edge $e$, which contains the information of the interaction along that edge. The partition function $Z_G$ is obtained by computing the overlap
\be\label{eq:ZGq}
Z_{G} = q\cdot \langle \psi_G^q|\alpha\ra.
\ee

\emph{Relation between $|\varphi\ra$ and $|\psi\ra$}.
Note that the state $|\psi_G\ra$ specializes to the state $|\varphi_G\ra$ if
the interaction between spins $s_a$ and $s_b$ connected by $e$ is such that $h_e(|s_{v_e^+} - s_{v_e^-}|_q) = 1$ if $|s_{v_e^+} - s_{v_e^-}|_q=0$, and $h_e(|s_{v_e^+} - s_{v_e^-}|_q) = 0$ otherwise, for all edges $e\in V$. It must also hold that the magnetic fields at every site $a$ are zero, i.e.~$h_a=0$, for all $a\in V$.

\begin{figure}[htb]
\centering
\psfrag{(a)}{\hspace{-1mm}$|\varphi_G\ra$}
\psfrag{(b)}{\hspace{-3mm}$|\psi_{G+h}\ra$}
\psfrag{v0}{$v_0$}
\includegraphics[width=0.5\columnwidth]{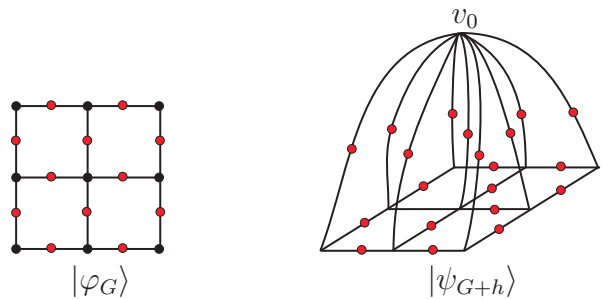}
\caption{
The state $|\varphi_G\ra$ is equivalent to the state $|\psi_{G+h}\ra$. The latter is obtained by taking the state $|\psi_G\ra$, adding a new vertex $v_0$ connected to all other vertices in $G$, and letting the interaction between $v_0$ and every vertex $a\in V$ equal the magnetic field at $a$.}
\label{fig:psi+spike}
\end{figure}

However, we can also incorporate magnetic fields in the description of $|\psi\ra$ in the following way. Consider the state $|\psi\ra$ defined on a certain graph $G=(V,E)$, $|\psi_G\ra$. Now add an extra vertex $v_0$ which is connected via decorated edges to all other vertices in $G$, thus forming the state denoted by $|\psi_{G+h}\ra$ (see figure~\ref{fig:psi+spike}). By letting $v_0$ interact with all other qubits with a coupling strength equal to the magnetic field, i.e.~$J_{0a}=h_a$ for all $a\in V$, the resulting state is $|\psi_{G+h}\ra = 2|\varphi_G\ra$ (see \cite{Hu08} for a proof). Note that throughout this paper we deal with non$-$normalized states.
Notice also that the number of particles in $|\varphi_G\ra$ and $|\psi_G\ra$ is the same: while $|\varphi_G\ra$ contains $|E|$ edge qubits and $|V|$ vertex qubits, $|\psi_{G+h}\ra$ contains $|E|+|V|$ edge qubits.

\subsection{Spin models in the GHZ picture}
\label{ssec:GHZpicture}

\begin{figure}[htb]
\centering
\psfrag{a}{$| GHZ_q \ra^{(a')}$}
\psfrag{b}{$|\chi_{i'}\ra$ }
\psfrag{q}{$q$}
\psfrag{d}{$d$}
\includegraphics[width=0.4\columnwidth]{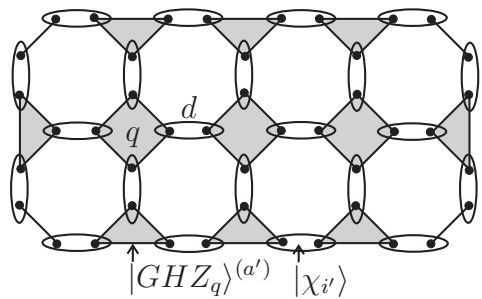}
\caption{GHZ picture of a $q-$state edge model on a 2D square lattice. Rhombi represent GHZ states of $q-$level systems, each of which involves $n_a$ (virtual) particles. Ellipses represent projections, which describe the interaction between different GHZ groups. Here all interactions are pairwise. Notice that the lines between the particles in the GHZ state have a different meaning than the edges between vertices in a graph state (which correspond to a controlled phase gate between two qubits).}
\label{fig:GHZ}
\end{figure}

Here we present the \emph{GHZ picture} of an edge model~\cite{Hu08}.
Consider first a $q-$state edge model on a 2D square lattice. We can represent it in its GHZ picture by substituting every classical particle sitting at a vertex by 4 (virtual) quantum particles in a GHZ state of $q-$levels: $|0000\ra+|1111\ra+\ldots+|q-1,q-1,q-1,q-1\ra$, and by substituting every interaction by a projection onto two quantum particles (each belonging to a different GHZ state) onto the state $|\chi_{ab}\ra=\sum_{s_a,s_b}e^{-\beta h(s_a,s_b)}|s_a s_b\ra$, where $h(s_a,s_b)$ is the part of the Hamilton function that describes the interaction between $s_a$ and $s_b$ (see figure~\ref{fig:GHZ}).

Then, the partition function of the edge model can be found by computing the overlap
\be
Z_{e,2D}=\la\chi|GHZ\ra
\ee
where $e$ indicates that it is an edge model, and $2D$ indicates that the interaction pattern is a 2D square lattice. Here $|\chi\ra$ is the tensor product of all the projections, $|\chi\ra = \bigotimes_{(a,b)\in E}|\chi_{ab}\ra$ (where $\{a,b\}$ are edges in the graph picture of the edge model), and $|GHZ\ra$ is the tensor product of the GHZ states associated with each particle, $|GHZ_q\ra= \bigotimes_{a}|GHZ_q\ra^{(a)}$.

For example, the GHZ picture of this edge model specializes to the Ising model without magnetic fields by setting $q=2$ and $|\chi_{ab} \rangle = e^{-\beta J_{ab}}(|00\ra + |11\ra) + e^{\beta J_{ab}}(| 01\ra + |10 \ra )$ for all edges $(a,b)$ in the 2D lattice.
Local magnetic fields can be incorporated in this picture by setting
$|\chi_{ab} \ra= \sum_{s_a,s_b=0}^1 \exp\{\beta [(-1)^{s_a+s_b}J_{ab} + (-1)^{s_a}h_a + (-1)^{s_b}h_b]\}|s_a s_b\ra$.
Similarly, the Potts model on a graph $G=(V,E)$ on $q-$level systems is obtained by setting
$|\chi_{ab}\rangle = e^{\beta J_{ab}}\sum_{j=0}^{q-1}|jj\rangle + \sum_{j\neq m}|jm\rangle$
 for all $(a,b)\in E$.

More generally, we can obtain the GHZ picture of a general class of edge models in the following way. Consider a $q-$state edge model in which spins interact according to the Hamilton function
\begin{equation}
\label{eq:hamiltonian-GHZpicture}
H(\mathbf{s})=-\sum_{i\in I}h_i (\mathbf{s}^{(i)}),
\end{equation}
where $i$ is an interaction (possibly involving many particles) within an index set $I$, $h_i$ is the Hamilton function associated with that interaction, and $\mathbf{s}^{(i)}$ are the spins involved in interaction $i$.

In order to obtain the GHZ picture of such an edge model~\cite{Hu08}, we replace each particle $a$ by $n_a$ (virtual) quantum particles in the state
\be
\sum_{i=0}^{q-1}|i\ra^{\otimes n_a}= |GHZ_q\ra^{(a)}.
\ee
Here $n_a$ is the number of interactions in which particle $a$ is involved (e.g., for a graph, $n_a$ is the number of adjacent edges it has).
The state formed by all particles is the product state of the GHZ state associated with each particle, viz.
\begin{equation}
|GHZ_q\ra=\bigotimes_{a\in V}|GHZ_q\ra^{(a)}.
\label{eq:GHZ-edge}
\end{equation}
 This is a stabilizer state, and the generators of its stabilizer group are $Z_1^{(a)}Z_{j}^{(a)}$ for $j \in\{2,\ldots,n_a\}$ and $\bigotimes_{j=1}^{n_a} X_j^{(a)}$ for all $a\in V$, where $X$ and $Z$ are the generalized Pauli operators of~\eref{eq:XZ}.

\begin{figure}[htb]
\centering
\includegraphics[width=0.35\columnwidth]{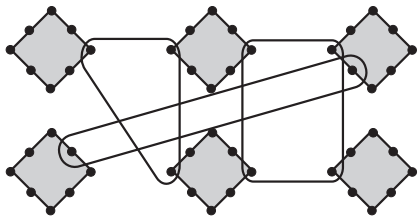}
\caption{A Hamilton function involving 2, 3 and 4$-$body interactions in the GHZ picture. Projections contain the information of the interaction pattern (which classical spin interacts with which, and whether many-body interactions are present), as well as the details of the interactions. Not all interactions are represented in the figure, that is why some quantum particles (black dots) are not projected.}
\label{fig:k-body-interaction}
\end{figure}

We also replace each interaction $i$ by a projection onto the state
\begin{equation}
\label{eq:chi}
|\chi_i\ra= \sum_{\mathbf{s}^{(i)}}e^{\beta h_i(\mathbf{s}^{(i)})}|\mathbf{s}^{(i)}\ra,
\end{equation}
which involves $|\{\mathbf{s}^{(i)}\}|$ quantum particles, each of which belongs to a different GHZ state (see figure~\ref{fig:k-body-interaction} for an example of how to depict many body interactions in the GHZ picture).
Note that the interaction pattern is encoded in these projections, since they specify which spin interacts with which. The dimension $d$ of the interaction particle corresponds to the number of different energy values that are assigned. Thus, for $k-$body interactions between $q$ level systems, the most general projection has dimension $d=q^k$.
Note that such projection includes all possible interactions between $2, \ldots, k-1$ bodies, as well as local terms. We define $|\chi\ra$ as the tensor product of all projections $|\chi\ra= \bigotimes_{i\in I} |\chi_i\ra$.
Then the partition function of such an edge model in the GHZ picture is found by computing
\be \label{eq:ZGq-GHZ}
Z_{e,I} = \la \chi |GHZ_q \ra.
\ee

\subsection{Spin models in the PEPS picture}
\label{ssec:PEPSpicture}

Here we present the \emph{PEPS picture} of a vertex model (PEPS stands for Projected Entangled Pairs)~\cite{Hu08,Ve04,Ve04a}. A vertex model represented on a graph $G=(V,E)$ is a model in which particles (possibly $q-$dimensional) are sitting at the \emph{edges} of the graph and vertices represent (typically many-body) interactions between particles. This is the Graph picture of a vertex model.

\begin{figure}[htb]
\centering
\psfrag{a}{$|\chi_v\ra$}
\psfrag{b}{$|\phi^{+}\ra$}
\includegraphics[width=0.4\columnwidth]{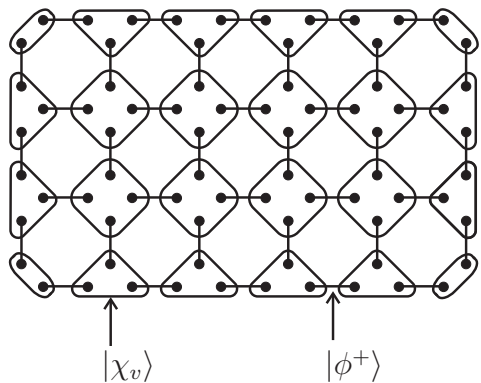}
\caption{PEPS picture of a vertex model on a 2D square lattice.}
\label{fig:peps}
\end{figure}

In order to obtain the PEPS picture of a vertex model, we proceed similarly as for the GHZ picture of an edge model. Consider first as an example a vertex model on a 2D square lattice. Its PEPS picture is obtained by substituting every particle $e$ by a Bell pair $\sum_{j=0}^{q-1} |jj\ra^{(e)}=: |\phi_q^+\ra^{(e)}$ (i.e.~a GHZ state of two particles), and every 4$-$body interaction by a projection $|\chi_v\ra$ onto one particle of each of the four Bell pairs that `end' in that vertex (see figure~\ref{fig:peps}).
Similarly, the state formed by all particles is the product state of the Bell state associated with each particle, viz.~$|\phi^+_q\ra=\bigotimes_{e\in E}|\phi_q^+\ra^{(e)}$, and we define $|\chi\ra$ as the product state of all interactions, $|\chi\ra =\bigotimes_{v\in V} |\chi_v\ra$
Thus, the partition function of this vertex model in the PEPS picture is obtained by computing
\be \label{eq:ZGq-Peps}
Z_{v,I} = \la \chi|\phi_q ^+ \ra.
\ee

For example, this vertex model can specialize to widely studied models such as the six-vertex model or the eight-vertex model. To do this, we express the 4 indices of a 4$-$body interaction as the 2 rows and 2 columns of a matrix, and we write $|\chi_v\ra$ as  $\chi_v = \sum_{ijkl}e^{\beta h(s_i, s_j, s_k, s_l)}|ij\ra\la kl|$.  Then this model specializes to the eight-vertex model by setting the function $h(s_i, s_j, s_k, s_l)$ such that
\begin{equation}
\chi_v=
\left(
\begin{array}{cccc}
e^{\beta E_1} & 0                    & 0                    & e^{\beta E_7} \\
0                   & e^{\beta E_2} & e^{\beta E_5} & 0                    \\
0                    & e^{\beta E_6} & e^{\beta E_3} & 0                    \\
e^{\beta E_8} & 0                   & 0                    & e^{\beta E_4}
\end{array}
\right),
\end{equation}
and the six-vertex model is obtained by additionally setting $e^{\beta E_7}=e^{\beta E_8}=0$~\cite{Bax82}.
This general class of  vertex models can also specialize to the topological colour codes introduced in~\cite{Bo08}, if we let the spins interact according to an hexagonal lattice, and we set the Bell states to have $q=2$ and the interactions $|\chi_v\ra = \sum_{s_a,s_b,s_c}e^{(-1)^{s_a+s_b+s_c}\beta J}|s_a+s_b+s_c\ra$ for all vertices $v$ in the lattice.

More generally, a PEPS picture can be seen as a specific case of a GHZ picture. That is, a picture of a spin model in which particles are substituted by GHZ states, and interactions by projections onto particles of different GHZ states. In the previous subsection we have shown how to obtain the GHZ picture of a general class of edge models. With the same procedure we would describe a general class of vertex models, that is, vertex models on $q-$level systems and with any combination of $1, 2,\ldots, k$ many-body interactions, for any $q$ and $k$, in the PEPS picture.

\section{Completeness of the 2D Ising model with complex parameters}\label{sec:completeness-2DIsing}

In this section we prove the first main result of this paper, that is, that \emph{the 2D Ising model is complete}, i.e.~that the partition function of any model of classical spins with any interaction pattern can be expressed as a special instance of the partition function of the classical Ising model with magnetic fields on an (enlarged) 2D square lattice. To prove it, we first show that it is complete for Ising models with magnetic fields on arbitrary graphs, then for edge models on $q-$level systems with $k-$body interactions (for any $q$ and $k$) and finally for vertex models on $q-$level systems with $k-$body interactions (for any $q$ and $k$ as well).
We also show that the enlargement of the 2D square lattice is polynomial for Ising models on all graphs, and for edge and vertex models with $q$ and $k$ bounded.
However, in order to obtain these results, one needs to allow for some coupling strengths and magnetic fields on the 2D Ising model to be complex, and thus not physical.

Finally, we will present another completeness result with complex parameters, namely that  \emph{the 2D vertex model with general interactions is complete with complex parameters for all edge and vertex models} (including Ising models on arbitrary graphs).

\subsection{Ising models with magnetic fields}
\label{ssec:completeness-Ising-complex}

To prove the completeness of the 2D Ising model for Ising models on arbitrary graphs, we make use of~\eref{eq:Z=alpha|phi} and of the knowledge of one-way quantum computation. The one-way quantum computer is a paradigm for quantum computation, where computations
are realized by performing single-qubit measurements on a
highly entangled substrate state called the 2D cluster
state $|{\cal C}\rangle$~\cite{Ra01b}; the latter is a
graph state~\cite{He06} associated with a 2D square lattice
${\cal C}$.

In particular, we use the fact that the one-way
quantum computer is \emph{universal}. This means that any $n-$qubit quantum state can be prepared, up to local unitary Pauli operations, by
performing sequences of single-qubit measurements on a
$d\times d$ cluster state $|{\cal C}\rangle$ of
sufficiently large system size $M=d^2$.
This implies
that every $n-$qubit quantum state $|\psi\rangle$ can be
written in the following way:
\be
\Sigma|\psi\rangle =2^{(M-n)/2} \left(I\otimes \langle\beta|\right)|{\cal
C}\rangle.
\label{psifromcluster}
\ee
This formula represents one measurement branch of a one-way
computation performed on an $M-$qubit cluster state,
yielding the state $|\psi\rangle$ (up to a local operation
$\Sigma$) as an output state on the subset of $n$ qubits which
have not been measured. The product state
$\langle\beta|=\bigotimes_j\langle\beta_j|$ only acts on the measured qubits, and is determined by the bases and
the outcomes of the different steps in the computation. The
local unitary operator $\Sigma$, called``correction
operator'', acts on the unmeasured qubits (i.e.~on the
Hilbert space of $|\psi\rangle$). The tensor factors of
$\Sigma$ are always instances of Pauli operators,
$\Sigma_i\in\{I, X, Y, Z\}$. The prefactor $2^{(M-n)/2}$
reflects the fact that the success probability of every
measurement branch is $2^{n-M}$.

The state $|\varphi_{2D}\rangle$ (i.e.~the state $|\varphi_{G}\rangle$ where $G$ is the
2D square lattice) is also a universal resource. This
is because the 2D cluster state $|{\cal C}\rangle$ can be
deterministically generated from $|\varphi_{2D}\rangle$ (up to a local correction) by performing
single-qubit $Y-$measurements on all edge qubits.
As a consequence,
one has
\be
\Sigma'|{\cal C}\rangle = 2^{|E|/2}\left(I\otimes \langle
0_Y|^{V_E}\right) |\varphi_{2D}\rangle,
\label{universality_decorated}
\ee
where $|0_Y\rangle^{V_E}$ is a tensor product of the $(+1)-$eigenstate of $Y$ on all edge qubits, and
$\Sigma'$ is a local correction.

\begin{figure}[htb]
\psfrag{a}{$\langle 0_Y |^{V_E}\quad$}
\psfrag{b}{$\langle \beta |$}
\centering
\begin{tabular}{ccccc}
\multicolumn{5}{c}{\includegraphics[width=0.7\columnwidth]{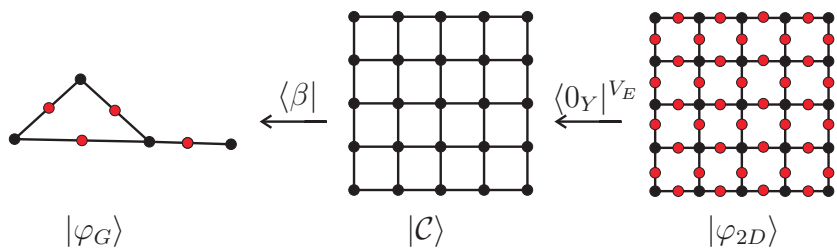}}\\
\hspace{.75cm}$|\varphi_{G} \rangle $ & &
\hspace{2.9cm}$|\mc{C}\rangle$ &&
\hspace{1.9cm}$|\varphi_{2D}\rangle$
\end{tabular}
\caption{The graph state $|\varphi_{G} \rangle$ (a specific instance of it is drawn in the figure) is generated from the cluster state $|\mc{C}\rangle$ by applying a measurement pattern $\langle \beta |$ on the cluster. In its turn, the cluster state is generated from the state $|\varphi_{2D}\rangle$ by measuring all edge qubits $V_E$ (denoted by red dots) in the $Y-$basis. Black dots denote vertex qubits. }
\label{fig:decorated-cluster->final-graph}
\end{figure}

We are now ready to establish the connection  between the
evaluation of Ising partition functions and universal MQC. To do so,
consider the following procedure.
First, the partition function of an Ising model on a graph $G$ can be expressed in the form~\eref{eq:Z=alpha|phi}.
The stabilizer state $|\varphi_{2D}\rangle$ is
obtained from the cluster state $|\mathcal{C}\ra$ after applying a certain
measurement pattern $|\beta \rangle$ (i.e.~we consider~\eref{psifromcluster}
with $|\psi\rangle = |\varphi_{2D}\rangle$).
Finally the cluster state $|\mathcal{C}\ra$ is obtained from the state $|\varphi_{2D}\ra$
after measuring all edge qubits in the $Y$ basis (see equation \eref{universality_decorated} and figure~\ref{fig:decorated-cluster->final-graph}).
This means that the partition function of the Ising model on the
graph $G$ can be written as
\be
Z_G(\{J_{ab}, h_a\})= A\: \langle \gamma | \varphi_{2D}\rangle,
\label{Z_G2D}
\ee
where $A=2^{(|E|+|V|+M-n)/2}$ is a constant and $|\gamma\rangle$ is a product state,
$|\gamma\rangle= \Sigma|\alpha\rangle \otimes
\Sigma'|\beta\rangle\otimes |0_Y\rangle^{V_E}. $
Now, by comparing the right hand side of~\eref{Z_G2D} with~\eref{eq:Z=alpha|phi}
we see that it corresponds to the partition function of the Ising model on a 2D square lattice evaluated with
a set of parameters $\{J_{ij}', h_i'\}$ determined by $|\gamma\rangle$.
This allows us to conclude that $Z_G$ can be written as
follows:
\be
Z_G(\{J_{ab}, h_a\})\propto
Z_{\mbox{\scriptsize{2D}}}(\{J_{ij}', h_i'\}).
\label{universality_ising}
\ee
In other words, the partition function of the Ising model on an
arbitrary graph can be recovered as a special instance of
the partition function of the Ising model on a 2D square lattice.
This proves that the 2D Ising model is complete for Ising models on any other graph. We now give a few remarks regarding this construction.

\emph{Interaction pattern and details of the interaction.}
Note that in~\eref{Z_G2D} the product state $|\gamma\rangle$
is determined by both the interaction graph $G$ and the
couplings $\{J_{ab}, h_a\}$ of the original model. For, on
the one hand, it contains the states $|\alpha_{ab}\rangle$
and $|\alpha_a\rangle$ encoding the couplings of the
original model; on the other hand, $|\gamma\rangle$
contains states $|\beta_j\rangle$ and $|0_Y\rangle$
corresponding to the sequence of one-qubit measurements
which are to be implemented in order to generate
$|\varphi_{G}\rangle$ from the universal resource
$|\varphi_{2D}\rangle$. In going from~\eref{Z_G2D} to~\eref{universality_ising}, the state
$|\gamma\rangle$ in turn determines the couplings with which
the 2D model is to be evaluated. Note that the decorated
cluster state $|\varphi_{2D}\rangle$ has
vertex qubits and edge qubits. The factors of
$|\gamma\rangle$ acting on the edge qubits determine the
pairwise interactions $J_{ij}'$, whereas the factors of
$|\gamma\rangle$ acting on the vertex qubits determine the
external fields $h_i'$. The tensor factors of
$|\gamma\rangle$ which act on the edge qubits are all equal
to $|0_Y\rangle \propto |0\rangle + i|1\rangle$. This
implies, in particular that in \eref{universality_ising}
only \emph{homogeneous} pairwise couplings $J_{ij}'$ need
to be considered. Hence, all information regarding the couplings $J_{ab}$, the external fields $h_a$ of
the original model and the graph $G$ of this model
will be encoded in the factors of $|\gamma\rangle$ acting
on the vertex qubits, and thus in the external fields
$h_i'$ (which will typically be inhomogeneous). This fact does no longer hold in our construction for the completeness results with real parameters (section~\ref{sec:completeness-real}).

\emph{Polynomial enlargement of the system size. }
We show that the number of qubits in the 2D square lattice
is polynomially enlarged w.r.t the number of qubits
in the final graph.
As proved in~\cite{Ra01}, for all $n-$qubit states
$|\psi\rangle$ that can be efficiently prepared in the
circuit model, i.e.~by a polynomial sequence
of  two-qubit gates, the required size $M$ of the cluster state
in~\eref{psifromcluster} scales polynomially with the
number of qubits: $M \propto {\rm poly}(n)$. Moreover, in
this case the measurement bases $|\beta_j\rangle$ as well
as the correction operations $\Sigma$ can be efficiently
determined.
Since any graph state on $n$ qubits can be prepared using
at most $O(n^2)$  controlled-phase gates~\cite{He06}, it
follows that an arbitrary $n-$qubit graph state
can be written in the form
\eref{psifromcluster} with $M=\mbox{poly}(n)$.
Furthermore, for the preparation of graph states every
single-qubit state $|\beta_j\rangle$ can always be chosen
to be one of the $X-$, $Y-$ and $Z-$eigenstates.
In conclusion, since $|\varphi_{G}\rangle$ is a stabilizer state, the
system size of the 2D cluster state grows polynomially with
the size of $G$. Furthermore, $|\beta\rangle$ consists of
$X-$, $Y-$ and $Z-$eigenstates.

\emph{Universality of triangular, hexagonal and Kagome lattices.}
Note that, in the above sequence of arguments, one step is particularly crucial, the universality of the 2D cluster states: this property is used to ``map'' an \emph{arbitrary} state  $|\varphi_{G}\rangle$, and hence the associated partition function, to the 2D cluster state, i.e~all states can be ``reduced'' to this single structure.
It was proven in~\cite{Va06a,Va06b} that not only the cluster state is universal, but graph states on a triangular, hexagonal and Kagome lattices are also universal resources. Therefore, the conclusions for the 2D square lattice apply also to any of these three lattices, i.e.~the Ising model on a 2D triangular, hexagonal or Kagome lattice is complete as well.
Also a 2D square lattice with holes has been proven to be universal, and hence the same completeness results hold for it as well~\cite{Br08}.

 \emph{Complex couplings and fields.}
In order to interpret measurements with couplings (or fields), one needs to identify the coefficients of the projection when expressed in the eigenbasis $\{|0\ra, |1\ra\}$ with the coefficients $e^{\beta J}$, $e^{-\beta J}$ (or $e^{\beta h}$, $e^{-\beta h}$), respectively. In this manner we unambiguously associate a projection with a coupling strength (or magnetic field).

However, the coefficients  $e^{\pm\beta J}$  are real and positive for all (real) values of $J$ (similarly for $h$). This means that if the coefficient of the projection is negative or complex, this will translate to a complex value of the coupling strength $J$ or the magnetic field $h$. For example, a projection of an edge qubit onto the eigenstate of $Y$ with eigenvalue +1, $|0_Y\ra \propto |0\ra + i|1\ra$ corresponds to a coupling strength $J_{ij}'= -i\pi/(4\beta)$ (or the same value of the magnetic field if applied on a vertex qubit). Note that we have used $Y$ measurements in the proof above in order to transform the state $|\varphi_{2D}\ra$ into $|\varphi_G\ra$, for a general $G$. More precisely, we have applied $Y$ measurements on every edge qubit to transform $|\varphi_{2D}\ra$ to the cluster state $|\mc{C}\ra$ (thus corresponding to a uniform, complex coupling strength between all vertex qubits, i.e.~$J_{ij}'=-i\pi/(4\beta)$ for all edges $(i,j)$ in the 2D lattice), and the measurements on the cluster state to transform it to $|\varphi_G\ra$ will generally involve $Y$ measurements as well (i.e.~some magnetic fields will take complex values $h_i'=-i\pi/(4\beta)$). This means that the 2D Ising model is complete \emph{with complex parameters} for Ising models on any other graph

Note that even though such complex coupling strengths or magnetic fields do not correspond to physical models, considering the partition function as a function with
complex arguments is commonly done, e.g., in the context of
evaluating the Tutte polynomial or finding
(complex) zeros of $Z_G$ to identify phase transition
points~\cite{Soxx}.

A simple analysis reveals when real parameters arise: within Pauli measurements, the projectors $|0\ra$, $|1\ra$ and $|+\ra = |0\ra+|1\ra$ correspond to real parameters, i.e.~$X$ and $Z$ measurements (recall that we only consider one branch of the outcomes); and in the Bloch sphere, projections onto $\cos(\theta/2) |0\ra + \sin(\theta/2) |1\ra$, with $\theta\in [0,\pi ]$ correspond to real parameters.
However, tilted measurements performed on a stabilizer state typically do not yield another stabilizer state~\cite{He06}, and our target state is a stabilizer state. Therefore, we shall  only consider Pauli measurements. This will be important in the next sections, where we address the question what completeness results can be obtained with real parameters.

\subsection{All edge models}
\label{ssec:completeness-complex-q}

Here we show that the partition function of the Ising model on a 2D square lattice can specialize to the partition function of a model in which the interacting particles are $q-$dimensional systems. Moreover, we prove that the 2D square lattice is polynomially enlarged w.r.t the graph on which the $q-$dimensional model is defined as long as $q$ is bounded. Even more generally, we allow for $k-$body interactions and we find that these models can also be generated efficiently provided $k$ and $q$ are bounded.
Even though an unbounded growth of $q$ or $k$ with the system size may require an exponential enlargement of the system size, we are able to give some examples in which either $q$ or $k$ is not bounded and we still obtain a polynomial enlargement of the system size.

\begin{figure}[htb]
\centering
\psfrag{a}{\hspace{-3mm}$m_q$}
\psfrag{r}{\hspace{-1mm}$m_q$}
\psfrag{2}{2}
\psfrag{b}{}
\psfrag{1}{$u_1$}
\psfrag{u2}{$u_2$}
\psfrag{3}{$u_3$}
\psfrag{n-1}{$u_{n-1}$}
\psfrag{n}{$u_{n}$}
\psfrag{c}{$m_d^{(i)}$}
\psfrag{d}{$U_i$}
\psfrag{q}{$q$}
\psfrag{x}{$\hspace{-2mm}|\chi_i\ra$}
\psfrag{e}{$\hspace{-2mm}|\chi_i\ra$}
\psfrag{=}{=}
\includegraphics[width=0.7\columnwidth]{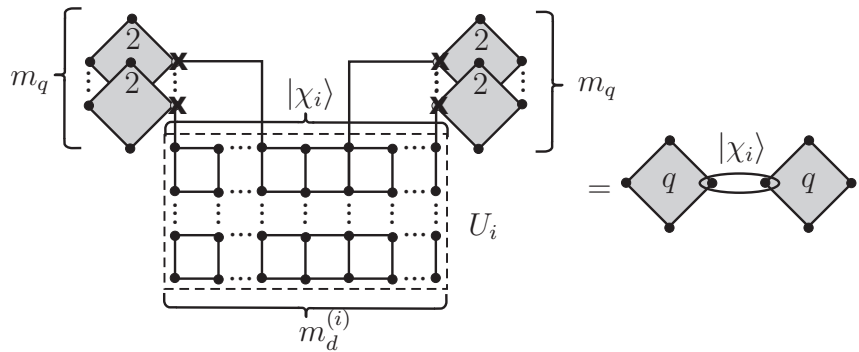}
\caption{A projection $|\chi_i\ra$  can be seen as a gate $U_i$ that prepares this state on the qubits involved in the interactions, and then these qubits are measured locally in the $X-$basis. For a projection of dimension $d$, the gate acts on $m_d^{(i)}=\lceil \log_2 d\rceil$ qubits,  $2m_d^{(i)}=2\lceil \log_2 d\rceil$ qubits in the one-way model. $q-$dimensional particles are viewed as $m_q$ two$-$dimensional particles. The gate $U_i$ can be decomposed into one$-$ and two$-$qubit gates, which correspond to a certain measurement pattern on the cluster. The dotted lines indicate variable size. }
\label{fig:GHZwithgate}
\end{figure}

\begin{figure}[htb]
\centering
\psfrag{a}{\hspace{-3mm}$m_q$}
\psfrag{r}{\hspace{5mm}$m_q$}
\psfrag{2}{2}
\psfrag{b}{}
\psfrag{1}{$u_1$}
\psfrag{u2}{$u_2$}
\psfrag{3}{$u_3$}
\psfrag{n-1}{$u_{n-1}$}
\psfrag{n}{$u_{n}$}
\psfrag{c}{\hspace{-3mm}$ m_d^{(i)}$}
\psfrag{d}{$U_i$}
\psfrag{chi}{$|\chi_i\ra$}
\psfrag{q}{$q$}
\psfrag{x}{$\hspace{-2mm}|\chi_i\ra$}
\psfrag{=}{=}
\includegraphics[width=0.7\columnwidth]{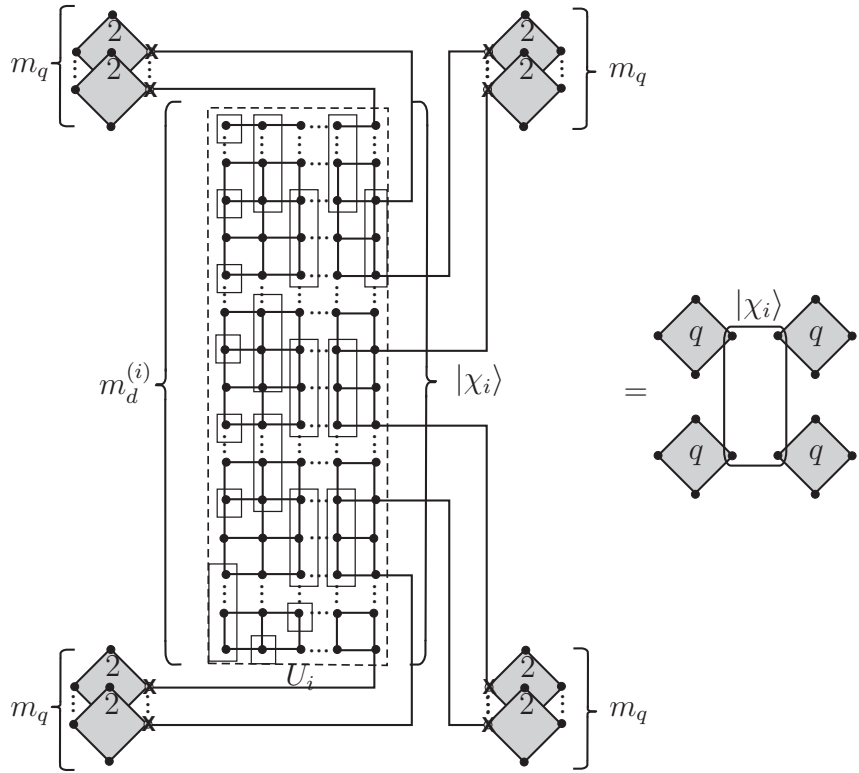}
\caption{A projection $|\chi_i\ra$ (of dimension $d$) can be seen as a gate $U_i$ acting on $m_d^{(i)}=\lceil \log_2 d\rceil$ qubits. $q-$level particles are viewed as $m_q$ two$-$level particles.}
\label{fig:GHZwithgate-4body}
\end{figure}

In order to prove these results, we use the GHZ picture presented in section~\ref{ssec:GHZpicture}. We want to express the partition function of~\eref{eq:ZGq-GHZ} as a special instance of the partition function of the 2D Ising model. To do so, we observe that the projection onto the state $|\chi_i\ra$ can be implemented by first preparing the state $|\chi_i\ra$ on the quantum particles on which the projection is to be performed (i.e.~qubits belonging to the different GHZ states), and then measuring locally these quantum particles in the $X$ basis.
The preparation of the state $|\chi_i\ra$ (of dimension $d$) can be done by a unitary gate $U_i$ acting on $m_d^{(i)}=\lceil \log_2 d^{(i)}\rceil$ qubits, $|\chi_{i}\ra = U_{i} |0\ra^{\otimes m_d^{(i)}}$ (see figure~\ref{fig:GHZwithgate} for a two$-$body interaction, and figure~\ref{fig:GHZwithgate-4body} for a 4$-$body interaction)).
Thus, the product state of all interactions can be expressed as $|\chi\ra = U|0\ra^{\otimes m_d}$
where $U = \bigotimes_{i\in I}U_i$ and $m_d=\sum_{i\in I}m_d^{(i)}$. This implies that the partition function of a $q-$state model can be written as
\be
Z_G^q = \la GHZ_q| U | 0 \ra^{\otimes m_d}.
\label{eq:completeness-qstate}
\ee
The state $|0 \ra^{\otimes m_d}$ is defined on qubits, and the state $ \la GHZ_q| U$ can be prepared from a cluster state by applying a certain measurement pattern on it. This proves the completeness of the 2D Ising model for $q-$state models.
Note that the preparation of the state $ \la GHZ_q| U$ will generally require measurements that contain the complex unit ``$i$'' or negative values, and thus the associated coupling strengths or magnetic fields in the 2D square lattice will be complex.

\emph{Efficiency results.}
We now study whether the 2D square lattice on which we compute the partition function has to be polynomially or exponentially enlarged w.r.t the graph on which the $q-$state model is defined.

We first observe that the state $| GHZ_q\ra$ can be efficiently prepared from a cluster state because it is a stabilizer state~\cite{He06}.
In fact, every $|GHZ_q\ra^{(a)}$ can be formed of $m_q=\lceil \log_2 q\rceil$ $|GHZ_2\ra$ states, simply by regarding the values of the latter as the binary expression of the $q$ levels that $|GHZ_q\ra$ can take (see figure~\ref{fig:GHZwithgate}). Taking into account that the state  $|GHZ_q\ra^{(a)}$ contains $n_a$ particles, the preparation of this state requires $\mathcal{O}(n_a m_q)$ qubits. Also the preparation of a GHZ state of $n$ particles requires $n-1$ phase gates. Hence the number of gates required to prepare $|GHZ_q\ra^{(a)}$ scales as $\mathcal{O}(n_a m_q)$.

\begin{figure}[htb]
\centering
\psfrag{a}{\hspace{-3mm}$m_q$}
\psfrag{z}{$\hspace{-1mm}m_q$}
\psfrag{2}{2}
\psfrag{b}{$|\chi_i\ra$}
\psfrag{mi}{\hspace{-3mm}$m_d^{(i)}$}
\psfrag{mj}{\hspace{-3mm}$m_d^{(j)}$}
\psfrag{c}{\vspace{3mm}\hspace{-3mm}$ m_d^{(i)}$}
\psfrag{d}{$U_i$}
\psfrag{e}{\hspace{-5.5mm}SWAPs}
\psfrag{f}{}
\psfrag{g}{$U_j$}
\psfrag{h}{$|\chi_j\ra$}
\psfrag{q}{$q$}
\psfrag{x}{$|\chi_j\ra$}
\psfrag{y}{$|\chi_i\ra$}
\psfrag{=}{=}
\includegraphics[width=0.7\columnwidth]{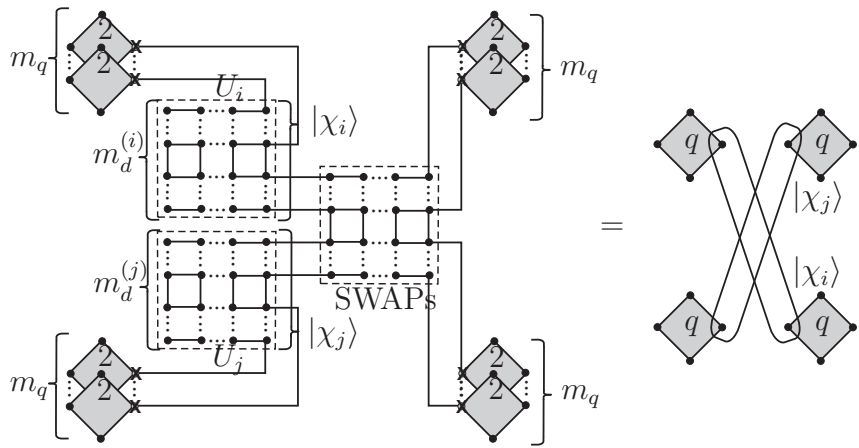}
\caption{A crossing of projections in the GHZ picture (right figure) is seen in the Graph picture as a the application of the gates corresponding to each projection followed by a SWAP gates that produce the crossing (left figure).
Auxiliary qubits are not represented in the gates for the sake of clarity.}
\label{fig:GHZwithgate-crossing}
\end{figure}

We now consider the efficiency in the generation of the unitary gates $U_i$.
Recall that each of these gates acts on $m_d^{(i)}=\lceil \log_2 d^{(i)} \rceil$ qubits initialized in $|0\ra$, where $d$ is at most $q^k$ for $k-$body interactions between $q-$level systems, i.e.~$m_d^{(i)} \leq k\log_2 q$. These gates can be implemented in the model of one-way quantum computation~\cite{Ra01}
(see figures~\ref{fig:GHZwithgate} and \ref{fig:GHZwithgate-4body}). Note also that whenever two projections cross each other, in the Graph picture this corresponds to implementating the gates specific to each projection and then applying several SWAP gates that exchange the position of the upper and lower lines and thus produce a crossing (see figure~\ref{fig:GHZwithgate-crossing}).  Any such gate can be decomposed into single and two qubit gates, each of which requires a constant number of qubits to be implemented in the one-way model.
Hence we need to study the number of single and two qubit gates required in the decomposition of $|\chi_i\ra$. In general an exponential number of gates might be required to prepare the state $|\chi_i\ra$. In the following we study which states $|\chi_i\ra$ require a polynomial number of gates.

Let us first assume that $q$ is bounded. A Hamilton function describing $k-$body interactions contains at most
\begin{equation}
\left(\begin{array}{c}n\\k\end{array}\right) \leq n^k
\end{equation}
 terms.
We have already noted that, if this $k-$body interaction has its maximum dimension $d=q^k$, this includes all possible $1, 2, \ldots, k-1$ interactions between these $k$ particles. If we further require that $k$ is bounded, then the unitary $U_i$ can be generated from a poly-sized circuit, and thus it requires a polynomial number of gates in the one-way model. It follows that any model with $q$ and $k$ bounded can be prepared efficiently.

If $k$ is unbounded, the number of terms in the Hamilton function may grow exponentially, and  the gate $U_i$ may be acting on an unbounded number of qubits. On the other hand, if $q$ is unbounded the gate $U_i$ may be acting on an unbounded number of qubits as well. Therefore, models with $k$ or $q$ unbounded will  generally require an exponential overhead in the system size.

We can find some exceptions to this result. In particular, we show that models in which $m_d$ remains bounded even though $q$ or $k$ grow unboundedly can still be generated efficiently. More precisely, we want to see that the Potts model with pairwise interactions~\cite{Wu84} but with $q$ unbounded can still be generated efficiently. The idea is that the Potts model only assigns two values of the energy ($J_e$ if they are in the same spin state, and 0 otherwise) independently of the number of spin states available $q$, and thus $m_q$.
The projection corresponding to this model is
\begin{eqnarray}
|\alpha_e\ra  =  (e^{\beta J}-1)(|0\ra|0\ra + |1\ra|1\ra + \ldots + \nonumber \\
|q-1\ra|q-1\ra) + \sum_{x,y=0}^{q-1} |x\ra |y\ra
 \end{eqnarray}
 for each edge $e\in E$.
Each $q-$level system is formed of $m_q=\lceil \log_2 q \rceil$ qubits, which we label $x_1\ldots x_{m_q}$ for one system and $y_1\ldots y_{m_q}$ for the other. Then the state $|\alpha_e\ra$ can be written as
\begin{equation}
|\alpha_e\ra =  (e^{\beta J}-1)\bigotimes_{i=1}^{m_q} |\phi_2^+\ra_{x_iy_i}+ |++\ra^{\otimes m_q},
\end{equation}
where $|\phi_2^+\ra_{x_iy_i}$ is a Bell pair between particles $x_i$ and $y_i$. Here $|+\ra$ and $|-\ra$ are the eigenstates of $\sigma_x$ with eigenvalues $+1$ and $-1$, respectively. By ordering the particles as $x_1 y_1 x_2 y_2 \ldots x_{m_q}y_{m_q}$, one can see that the maximum rank of the reduced density matrix of all bipartitions between the first $k$ particles and the rest of the system is 3.
This means that, in the Matrix Product State (MPS) representation of this state, all tensors have dimension 3 independent of the number of qubits $m_q$~\cite{Vi03}. Since this representation is efficient, the state can be efficiently generated~\cite{Sch05}.

\subsection{All vertex models}
\label{ssec:2DIsingforvertex}

We want to show that the partition function of the 2D Ising model can also specialize to the partition function of vertex models with $q-$dimensional particles with an arbitrary interacion pattern.

\begin{figure}[htb]
\centering
\psfrag{a}{$\hspace{-2mm}m_q$}
\psfrag{r}{$m_q$}
\psfrag{b}{$|\chi_i\ra$}
\psfrag{2}{2}
\psfrag{q}{$q$}
\psfrag{x}{$|\chi_i\ra$}
\psfrag{e}{$U_i$}
\psfrag{d}{$\hspace{-3mm}m_d^{(i)}$}
\psfrag{=}{=}
\includegraphics[width=0.6\columnwidth]{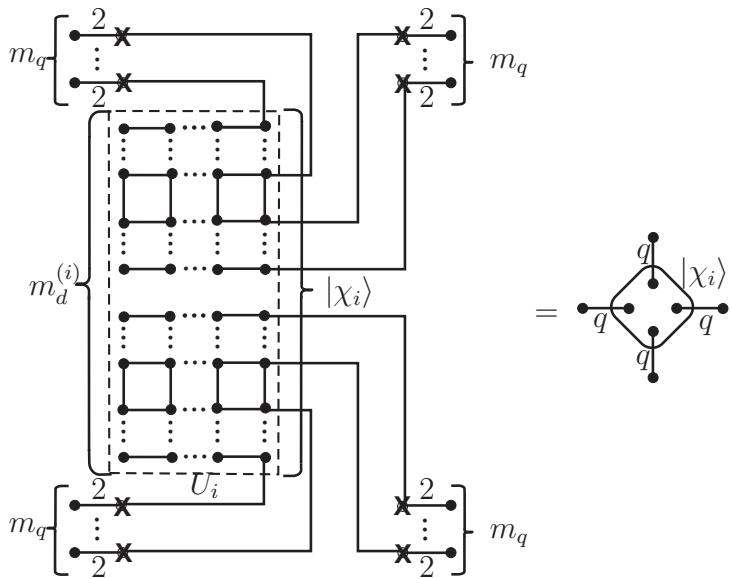}
\caption{A projection onto a state $|\chi_i\ra$ can be seen as a preparation of this state by the gate $U_i$ and the measurement of the particles in the $X-$basis. }
\label{fig:pepswithgate-4body}
\end{figure}

In section~\ref{ssec:PEPSpicture} we presented the PEPS picture of a vertex model, and we noted that it is a particular case of the GHZ picture. It follows that the completeness proof is completely analogous as for edge models in the GHZ picture.
We conclude that the partition function of a vertex model on $q-$level systems can be written as
\be
Z_{G,v}^q = \la \phi^+_q| U | 0 \ra^{\otimes m_d},
\label{eq:completeness-vertex}
\ee
from where the state $|0 \ra^{\otimes m_d}$ is defined on qubits, and the state $ \la \phi^+_q| U$ can be prepared from a cluster state by applying a certain measurement pattern on it (see figure~\ref{fig:pepswithgate-4body} for a 4$-$body interaction). This proves the completeness of the 2D Ising model for vertex models on $q-$level systems, for arbitrary $q$ and with an arbitrary interaction pattern.
Note that the preparation of the state $ \la \phi^+_q| U$ will also generally require measurements that contain the complex unit ``$i$'' or negative values, and thus the associated parameters will lie in the complex regime.

\emph{Efficiency results.}
Since the PEPS picture is a special case of the GHZ picture, it follows that the efficiency results are the same. That is,
whenever $q$ and $k$ are bounded, the 2D square lattice requires a polynomial enlargement w.r.t to the graph where the vertex model is defined. A model with $q$ or $k$ unbounded may require in general an exponential overhead.

\subsection{Completeness of the 2D vertex model with complex parameters} \label{sec:completeness-2D2vertex}

Here we want to show that a vertex model with two$-$level systems ($q=2$) defined on a 2D square lattice and with one-dimensional four body interactions is complete with complex parameters for the same models for which the 2D Ising model is complete.

The proof is based on the following observation: the PEPS picture of a vertex model on a 2D square lattice with two$-$level systems with one$-$dimensional projections (as the one depicted in figure~\ref{fig:peps}) corresponds indeed to the PEPS picture of a 2D cluster state~\cite{Ve04,Ve04a}.
More precisely, a single qubit measurement in a 2D cluster state would correspond in the PEPS picture to one-dimensional projections with a Hadamard rotation on two of the four qubits, since the edges in a cluster state correspond to the states $|0+\ra + |1-\ra$, whereas the pairs (which now act as edges) are in the state $|00\ra + |11\ra$ (note that $|\pm\ra \propto |0\ra \pm |1\ra$).
This means that all states that can be prepared from a cluster state (with measurements corresponding to complex parameters) can also be prepared from a 2D two$-$state vertex model (with complex parameters as well). Since the 2D cluster state is universal, it follows that a 2D two$-$state vertex model is complete with complex parameters for all $q-$state edge and vertex models. Note that all efficiency results for the 2D Ising model also hold for this 2D two$-$state vertex model.

\section{Example: Partition function of the 3D Ising model obtained from the partition function of the 2D Ising model} \label{ex:2Dto3D}

As an explicit example of the construction presented above, we show how to compute the partition of the Ising model with magnetic fields on a 3D square lattice (3D Ising model) by computing the partition function of the Ising model on a (polynomially) enlarged 2D square lattice. To do so, we only need to find a measurement pattern $\langle \beta |$ such that, when applied on the state $|\varphi_{2D} \ra$, the resulting state is $|\varphi_{3D}\ra$, i.e.~the stabilizer state $|\varphi\ra$ defined on a 3D square lattice.
In principle we could generate $|\varphi_{3D}\ra$ by implementing gates (consisting of measurements) on the cluster state. We present here a more direct procedure that requires a smaller cluster to generate the same $|\varphi_{3D}\ra$, and that is translationally invariant (if the initial resource state is $|\varphi\ra$ on a 2D square lattice with crossings, instead of $|\varphi_{2D}\ra$).

This procedure is based on measurement rules, which translate into a certain measurement pattern $\la \gamma |$ applied on the initial state defined on a certain graph $G$ (e.g. $|\varphi_G\ra$), transforming it to the same state defined on another graph $G'$ (e.g. $|\varphi_{G'}\ra$, with $|\varphi_{G'})\ra=(I\otimes \la \gamma|)|\varphi_G\ra$). It is clear from this definition that the measurement rules depend on the state they act upon.

Here we will make use of the following measurement rules for the state $|\varphi\ra$ (see~\eref{eq:varphi} for the definition of $|\varphi\ra$):
\begin{eqnarray}
(\la +_{a'}| \la 0_{a'b'}| \otimes I ) |\varphi_G\ra = |\varphi_{G'}\ra &\quad &\textrm{(merge rule)}
\label{eq:mergerule}\\
(\la +_{a'b'}| \otimes  I ) |\varphi_G\ra = |\varphi_{G''}\ra &\quad& \textrm{(deletion rule)}
\label{eq:deletionrule}\\
(\la \gamma| \otimes  I ) |\varphi_G\ra = |\varphi_{G'''}\ra &\quad& \textrm{(crossing)}
\label{eq:creationofacrossing}
\end{eqnarray}
where $G'$ is the same graph as $G$ but with the edge $a'b'$ merged (or: ``contracted''), $G''$ is the same as $G$ but with the edge $a'b'$ deleted, and $G'''$ is the same as $G$ but in which two pairs of particles now have a crossed interaction and other particles have been deleted as specified below.

\begin{figure}[htb]
\centering
\psfrag{=}{=}
\psfrag{(a)}{(a)}
\psfrag{(b)}{(b)}
\psfrag{(c)}{(c)}
\psfrag{(d)}{(d)}
\includegraphics[width=0.4\columnwidth]{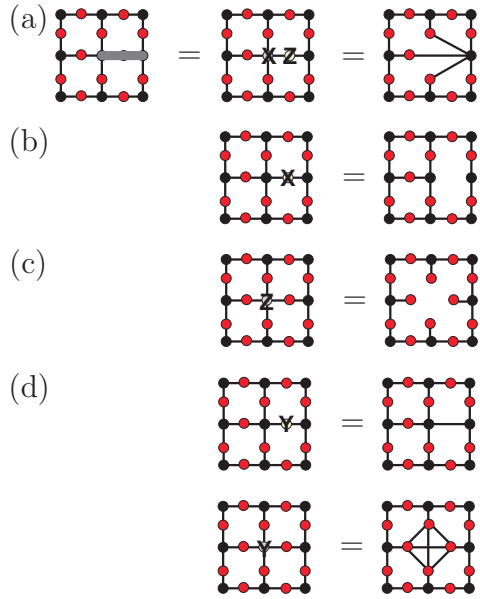}
\caption{Measurement rules for the state $|\varphi\ra$. (a) Merge rule. We denote the three qubits that are merged with a thick grey line on them. First, the $X$ measurement on the vertex qubit is performed, and then the $Z$ measurement on the chosen neighbor. (b) Deletion rule. An edge qubit is deleted by applying an $X$ measurement on it. (c) Deletion of a vertex qubit by a $Z$ measurement. (d) $Y-$measurement on an edge qubit and on a vertex qubit. Only rules (a) and (b) result in a decorated graph. Measurements (a), (b) and (c) correspond to real parameters, measurements (d) correspond to complex parameters.}
\label{fig:setofrules}
\end{figure}

\begin{figure}[htb]
\centering
\psfrag{(a)}{(a)}
\psfrag{(b)}{(b)}
\psfrag{=}{=}
\includegraphics[width=0.45\columnwidth]{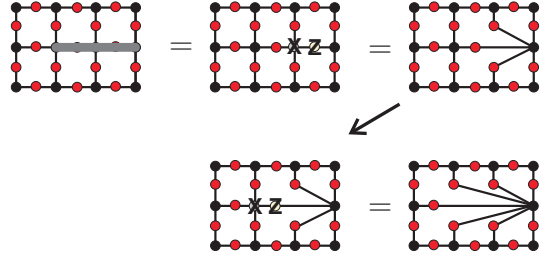}
\caption{Concatenation of merge rules. A thick gray line over certain vertex qubits  (and edge qubits inbetween) symbolize that they have to be merged. }
\label{fig:doublemergingrule}
\end{figure}

A merge rule is a measurement pattern according to which two vertex qubits and the edge qubit in between are merged into one vertex qubit. The neighbors of the resulting vertex qubit are the neighbors of the two vertex qubits that have been merged (see figure~\ref{fig:setofrules}(a)).
Several merge rules can be concatenated, thus merging several vertex qubits and the edge qubits inbetween into one vertex qubit (see figure~\ref{fig:doublemergingrule}). According to section~\ref{ssec:completeness-Ising-complex}, we can associate each projection with a value of the coupling strength or magnetic field, see equations~\eref{eq:alpha-phi-a} and \eref{eq:alpha-phi-ab}. In this way we see that the projection of the vertex qubit $a'$ onto $\la +|$ corresponds to setting the magnetic field $h_{a'}=0$, and the projection of the edge qubit $a'b'$ onto $\la 0|$ corresponds to setting the coupling strength $J_{a'b'}=\infty$.
 That is, an infinite coupling strength between two particles effectively corresponds to having one particle. Interestingly, we also see that this rule only requires real parameters to be realized. Whether a rule requires real or complex parameters will be a relevant issue for the completeness results with real parameters that we will present in section~\ref{sec:completeness-real}.

We would like to make a side remark here. We have just seen that one requires to set $J_{a'b'}=\infty$ in the above-presented rule. One may notice that, when this limit is taken, the state
$e^{\beta J_{a'b'}}|0\rangle + e^{-\beta J_{a'b'}}|1\rangle \approx
e^{\beta J_{a'b'}}|0\rangle$ instead of the state $|0\rangle$ is obtained, which may seem problematic because of the diverging normalizing factor. Here we note that there is a formulation of the present results in which this problem does not occur. This involves a rescaled
version of the Ising model, considering an interaction which assigns a value
$J_{ab}$ to non-aligned spins and the value 0 to aligned spins. While being equivalent to the  formulation of the Ising Hamilton function \eref{eq:H-phi},  in this formulation the states
(\ref{eq:alpha-phi-ab}) become $|0\rangle +  e^{-\beta J_{ab}}|1\rangle$. Hence the state $|0\rangle$ is also obtained by taking the limit $J_{ab}\to\infty$, but without the occurrence of a very large constant. It is easy to verify that all results in the present paper carry through completely in this rescaled formulation. Thus we shall henceforth stick to the conventional formulation of the Ising Hamilton function \eref{eq:H-phi}.

A deletion rule is a measurement pattern according to which two vertex qubits become disconnected (see figure~\ref{fig:setofrules}(b)).
A projection of an edge qubit $a'b'$ onto $\la + | = \la 0|+\la 1|$ corresponds to a coupling strength $J_{a'b'}=0$, that is, qubits $a'$ and $b'$ are not coupled to each other, which corresponds to having no edge between $a$ and $b$. We see, thus, that a deletion rule is realized with real parameters as well.

\begin{figure}[htb]
\centering
\psfrag{=}{=}
\includegraphics[width=0.4\columnwidth]{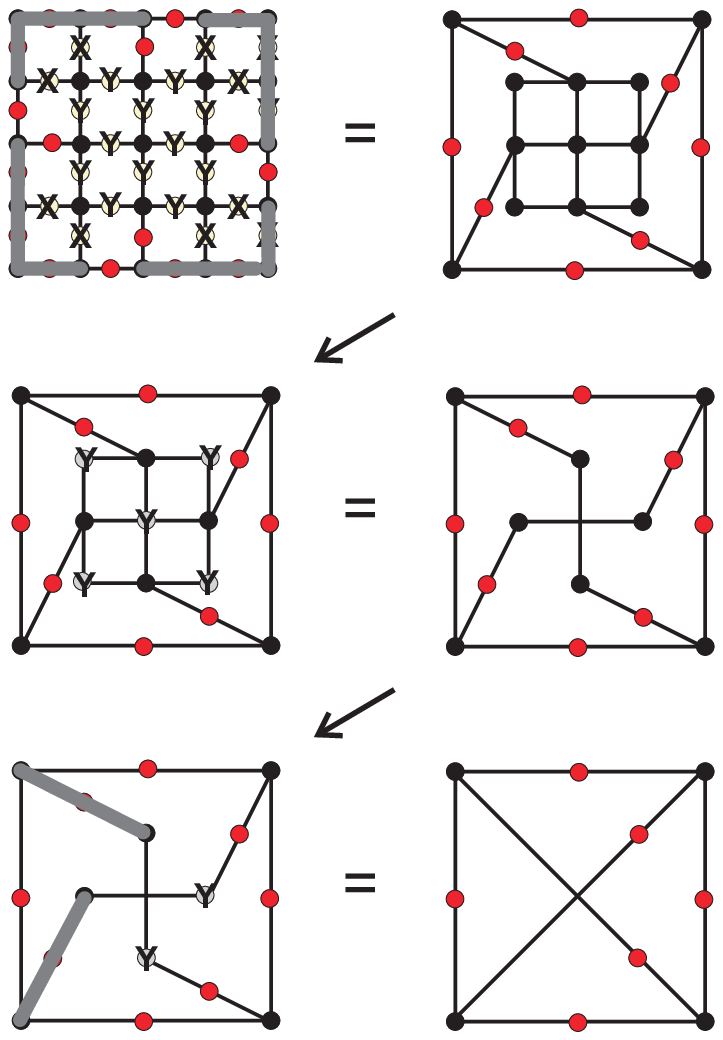}
\caption{
Measurement pattern to obtain a plaquette with a decorated crossing starting from a 2D square lattice. It involves $Y$ measurements, and hence it corresponds to complex parameters. The figure shows explicitly how the underlying graph changes with the measurements applied.}
\label{fig:decoratedcrossing}
\end{figure}

The creation of a crossed interaction between two pairs of particles corresponds to creating a crossing of edges. This requires the measurement pattern $\la \gamma |$ that is specified in figure~\ref{fig:decoratedcrossing}, which involves $Y$-measurements, and hence corresponds to complex parameters.

In the following we shall prove these three rules. Our strategy will be to consider the measurement rules known for graph states (see, e.g.,~\cite{He06}) and translate them to the state $|\varphi\ra$ (in~\cite{Hu08} the modification of the state is studied by looking at the stabilizer state). We recall that the state $|\varphi_G\ra$ can be obtained from the graph state defined on the decorated version of $G$, $|\tilde{G}\ra$, by applying a Hadamard rotation on all edge qubits, $|\varphi_G\ra = \bigotimes_{ab} H_{ab} |\tilde{G}\ra$.

The reader may skip these proofs and proceed directly with the example.

\emph{Proof of the merge rule}.
The merge rule for graph states reads: first $X-$measure the vertex qubit $a$ and then $Z-$measure the chosen neighbor $b_0$~\cite{He06}. The $X$ measurement requires the correction operator
\begin{equation}
\sqrt{i Y}= \frac{i}{\sqrt{2}}
\left[ \begin{array}{cc}
1&-1\\
1&1
\end{array} \right]
\end{equation}
on the chosen neighbor. The chosen neighbor is the edge qubit on which the $Z$ measurement is to be performed. This means that a projection of the chosen neighbor onto $\la 0 |$ actually becomes $\la 0 |(\sqrt{iY})^\dagger \propto \la +|$, i.e.~a projection onto an eigenstate of $X$.
The $X$ measurement also requires the correction operators $Z$ on the neighbors of $a$ which are neither $b_0$ nor neighbors of $b_0$.

We now translate the merge rule for graph states to a merge rule for the state $|\varphi\ra$.
The $X$ measurement on the vertex qubit remains the same for $|\varphi\ra$. The projection of the chosen neighbor onto $\la +|$ becomes $\la 0|$, i.e.~a $Z$ measurement, after the Hadamard rotation. The correction operators $Z$ on the other neighbors of $a$ become $X$ operators.

Note that the merge rule (the measurements and the corrections) all correspond to real parameters.
More precisely, first we regarded projections onto $\la 0|$, $\la 1|$ and $\la 0|+\la 1|$ as corresponding to real parameters. Now we see that, when taking the corrections into account, only the projections $\la 0|$ and $\la 0|+\la 1|$ correspond to real parameters.

\emph{Proof of the deletion rule.}
In the state $|\varphi\ra$ the interaction between two particles $a$ and $b$ (the original classical spins) is described by the edge particle $ab$. Hence, a deletion rule corresponds to deleting the edge particle and also the edges between particles $a$ and $b$.

To delete a particle in a graph state one has to measure its spin in the $Z-$ direction ($Z-$measure it). This measurement (more precisely, a projection on $\la 0|$) requires no correction operators on its neighboring qubits. Thus the $Z$ measurement becomes an $X$ measurement on an edge qubit in the state $|\varphi\ra$.

\emph{Proof of the creation of a crossing.}
In order to show how to create a crossing, we need to introduce two measurement rules: the \emph{deletion of a vertex qubit}, and \emph{$Y-$measurements} on edge and vertex particles.
The deletion of a vertex qubit consists of measuring a vertex qubit in such a way that this qubit becomes disconnected from all its neighboring edge qubits.
As pointed out above, the deletion rule in a graph state consists of a $Z$ measurement, and it requires no correction. Hence, a $Z$ measurement on a vertex qubit of $|\varphi\ra$ corresponds to deleting that qubit (see figure~\ref{fig:setofrules}(c)).
Regarding $Y$ measurements, for graph states it is known that a $Y$ measurement applied on a particle transforms the graph state by applying the local complementation rule to that particle (i.e.~deleting or adding edges between the neighbors of that particle if the edges were or were not there, respectively; see~\cite{He06}) (see figure~\ref{fig:setofrules}(d)).

We combine these two rules in the measurement pattern of figure~\ref{fig:decoratedcrossing} in order to create a decorated crossing (that is, the edges that cross each other are also decorated).
This concludes the proofs of the measurement rules.

\begin{figure}[htb]
\centering
\psfrag{=}{=}
\psfrag{a}{\hspace{-6.5mm}\small{complex}}
\psfrag{c}{\hspace{-5mm}\small{param.}}
\psfrag{b}{\hspace{-3mm}\small{real}}
\psfrag{d}{\hspace{-5mm}\small{param.}}
\psfrag{(a)}{\hspace{-2mm}\small{$|\varphi_{2D}\ra$}}
\psfrag{(b)}{}
\psfrag{(c)}{\hspace{-3mm}\small{$|\varphi_{3D}\ra$}}
\includegraphics[width=0.6\columnwidth]{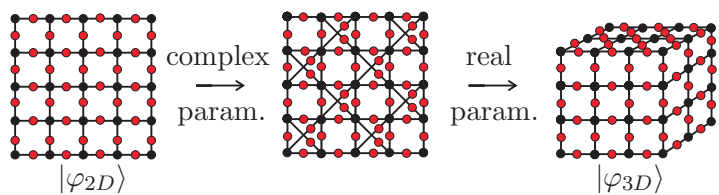}
\caption{We will first obtain a decorated 2D square lattice with crossings from a decorated 2D square lattice. This procedure involves complex parameters. From the latter lattice, we will obtain a decorated 3D square lattice, in a procedure which involves only real parameters.}
\label{fig:summary-example}
\end{figure}

Now we can proceed with the example. The outline of the procedure is as follows: first, we generate a 2D square lattice with crossings, with all edges decorated, from a 2D decorated square lattice. This involves creation of crossings, which correspond to complex parameters. Then, we generate a decorated 3D square lattice from the decorated 2D square lattice with crossings. This involves only merge and deletion rules, which correspond to real parameters (see figure~\ref{fig:summary-example}).

\begin{figure}[htb]
\centering
\includegraphics[width=0.40\columnwidth]{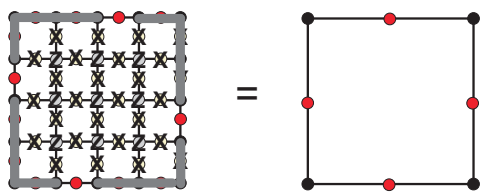}
\caption{Measurement pattern to generate a plaquette without a crossing. The plaquette needs to have the same size as the plaquette with a crossing. This pattern consists of deleting all inner vertices and merging the edges at the boundaries so that only one decoration remains at each side.}
\label{fig:decoratedempty}
\end{figure}

In order to generate a decorated 2D square lattice with crossings from a decorated 2D square lattice, we need to generate plaquettes with decorated crossings, and plaquettes without crossings. To obtain a plaquette with a crossing we proceed as in figure~\ref{fig:decoratedcrossing}. To obtain a plaquette without a crossing, we only need to select a square in the 2D square lattice of the same size as in figure~\ref{fig:decoratedcrossing}, and delete all vertices inside. Then we merge the vertices at the boundaries so that only one decoration remains at each side of the square (see figure~\ref{fig:decoratedempty}).

\begin{figure}[htb]
\psfrag{(a)}{(a)}\psfrag{(b)}{(b)}
\psfrag{a}{\hspace{-1mm}a}
\psfrag{1}{a'}
\psfrag{A}{a''}
\psfrag{b}{\hspace{-1mm}b}
\psfrag{2}{b'}
\psfrag{B}{b''}
\psfrag{c}{\hspace{-1mm}c}
\psfrag{3}{c'}
\psfrag{C}{c''}
\psfrag{d}{\hspace{-1mm}d}
\psfrag{4}{d'}
\psfrag{D}{d''}
\psfrag{e}{\hspace{-1mm}e}
\psfrag{5}{e'}
\psfrag{E}{e''}
\psfrag{f}{\hspace{-1mm}f}
\psfrag{6}{f'}
\psfrag{F}{f''}
\psfrag{g}{\hspace{-1mm}g}
\psfrag{7}{g'}
\psfrag{G}{g''}
\psfrag{h}{\hspace{-1mm}h}
\psfrag{8}{h'}
\psfrag{i}{\hspace{-1mm}i}
\centering
\includegraphics[width=0.65\columnwidth]{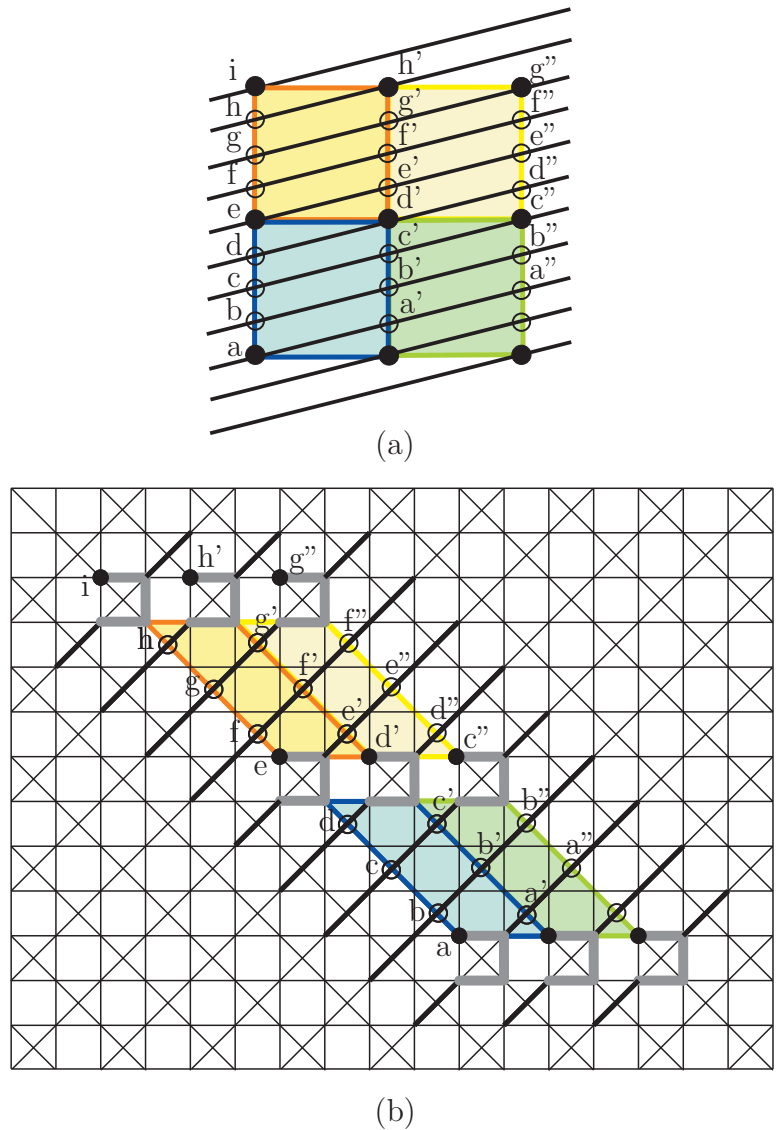}
\caption{(a) Face of 4 squares. (b) Embedding of this face on a 2D lattice with crossings. A square of a certain colour in (a) corresponds to the square of the same colour in (b). All crossings are indicated with an empty circle. Most vertices and crossings are labelled with a letter to indicate the correspondence of points between (a) and (b). All edges in (a) and (b) are decorated, even though this is not represented. Wide gray lines indicate vertices that have to be merged.
 The edges of the underlying 2D lattice with crossings that have no other line on top of them have to be deleted by deleting its decoration.
 Finally, edges $a$ to $d$ have to be merged, so that only one decoration remains between $a$ to $e$, and analogously for other sides of the squares.}
\label{fig:4colourfulsquares+in2Dlattice-colorsof6squares}
\end{figure}

Now we want to generate a 3D square lattice starting from a 2D square lattice with crossings by means of the merge and deletion rule alone. To do so, we first embed the figure shown in figure~\ref{fig:4colourfulsquares+in2Dlattice-colorsof6squares}(a), which we call a ``face'', on the 2D square lattice with crossings (the face can be seen as part of a three-dimensional structure, as will be made explicit later in figure~\ref{fig:3D-3Dprojected2D-allcolors}). We do it by tilting every square to the left so that the former vertical lines of the squares now coincide with the diagonal lines (going from the upper left corner to the lower right corner) of the 2D lattice with crossings (coloured tilted squares in figure~\ref{fig:4colourfulsquares+in2Dlattice-colorsof6squares}(b)). Only the lines crossing the squares remain to be embedded, and these are identified with the diagonal lines going from the lower left corner to the upper right corner in the 2D square lattice with crossings (thick black lines in figure~\ref{fig:4colourfulsquares+in2Dlattice-colorsof6squares}(b)). As shown in the figure, some vertices have to be merged, and also the sides have to be merged so that only one decoration remains at each side. Finally we apply the deletion rule to all edges that do not correspond to any edge of the projected 3D square lattice.

Note that the embedding of larger faces can be straightforwardly derived from this one. For example, an addition of squares in the horizontal direction in figure~\ref{fig:4colourfulsquares+in2Dlattice-colorsof6squares}(a) would translate to an addition of adjacent tilted squares in (b). Likewise, an addition of squares in the vertical direction would translate to an addition of ``layers of tilted squares'' in (b).

\begin{figure}[htb]
\psfrag{(a)}{(a)}
\psfrag{(b)}{(b)}
\centering
\includegraphics[width=0.6\columnwidth]{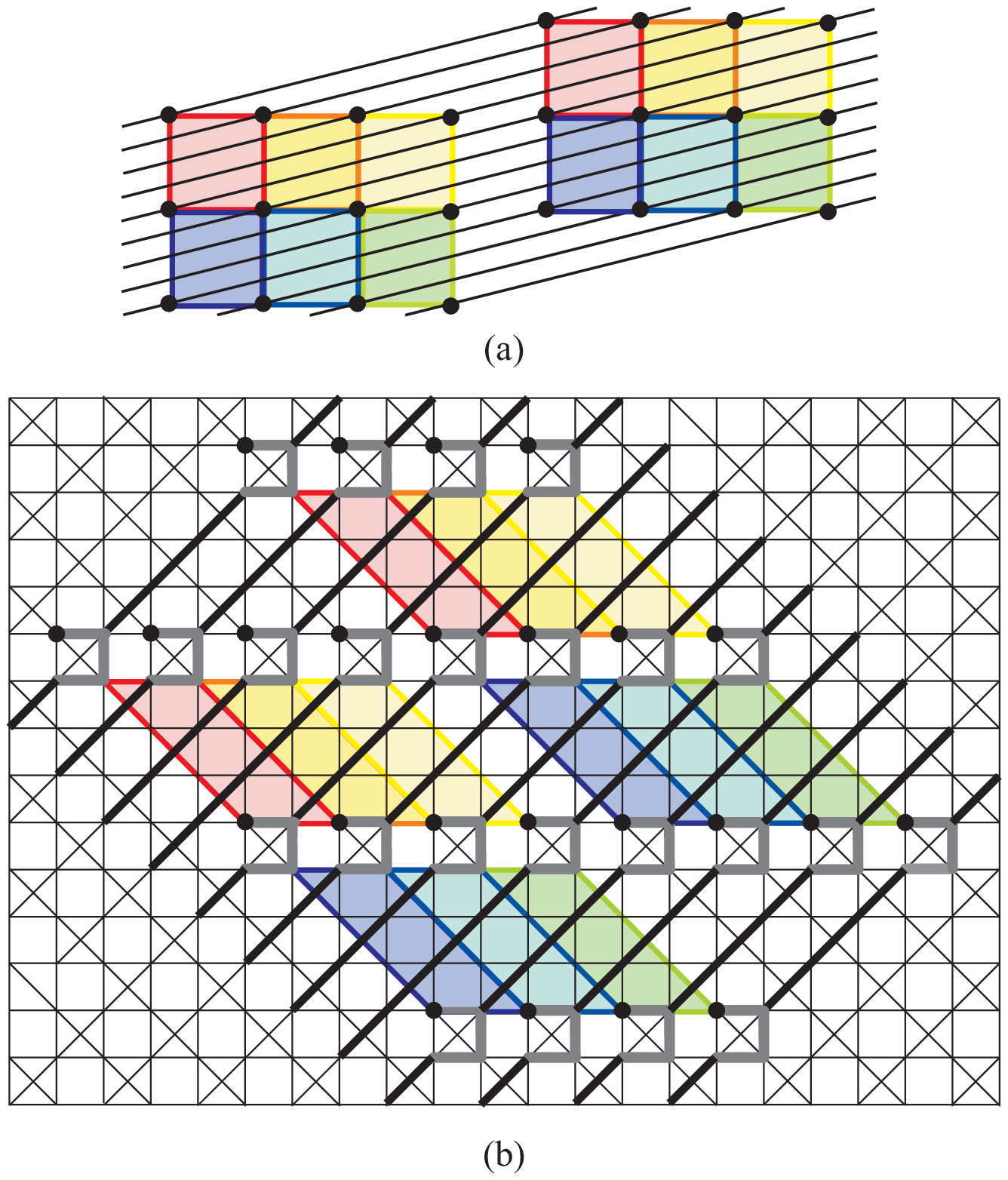}
\caption{(a) 2 concatenated faces of 6 squares each. (b) Embedding of these two faces in a 2D square lattice with crossings. The remarks of figure~\ref{fig:4colourfulsquares+in2Dlattice-colorsof6squares} hold for this one as well. This embedding is translationally invariant if periodic boundary conditions are considered.}
\label{fig:concatenation-embedding-6squares}
\end{figure}

Now we embed the concatenation of faces shown in figure~\ref{fig:concatenation-embedding-6squares}(a) in the 2D square lattice with crossings. To do so, we shift the position of one embedded face down and to the left w.r.t the other one (figure~\ref{fig:concatenation-embedding-6squares}(b)). In this way the lines that join one face with the other naturally follow their way while crossing the squares appropriately.  Note that this embedding is translationally invariant if periodic boundary conditions are considered.

\begin{figure}[htb]
\psfrag{(a)}{(a)}
\psfrag{(b)}{(b)}
\psfrag{(c)}{(c)}
\centering
\includegraphics[width=0.6\columnwidth]{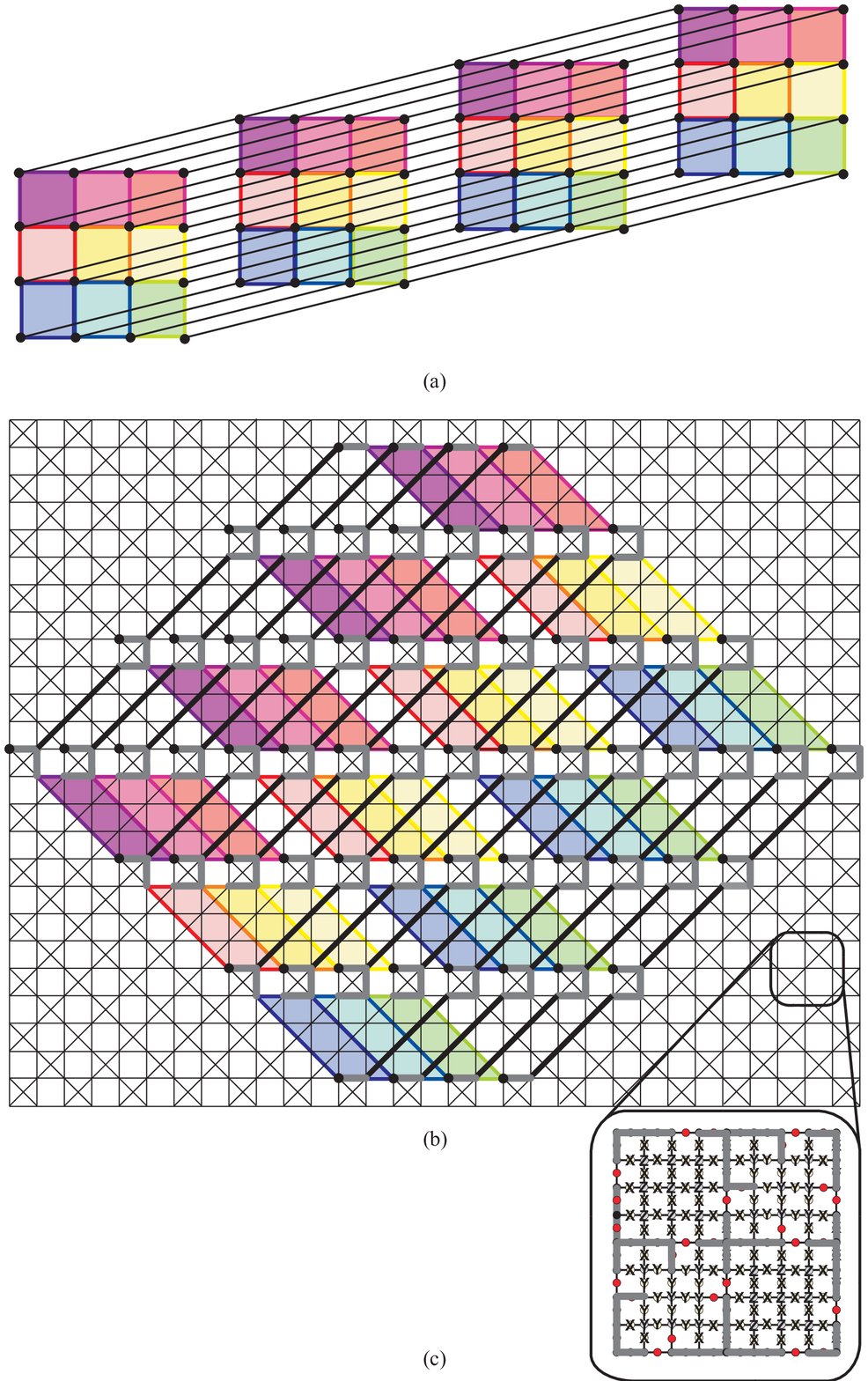}
\caption{(a) 4 concatenated faces, each of them with 9 squares, with open boundary conditions. (b) Embedding of this figure in a decorated 2D square lattice with crossings. (c) The decorated 2D square lattice with crossings is prepared from a decorated 2D square lattice by implementing the measurement pattern specified in the figure. This summarizes the procedure.}
\label{fig:concatenation-embedding-obc}
\end{figure}

\begin{figure}[htb]
\psfrag{x}{$x$}
\psfrag{y}{$y$}
\psfrag{z}{$z$}
\centering
\begin{tabular}{cc}
\multicolumn{2}{c}{\includegraphics[width=0.65\columnwidth]{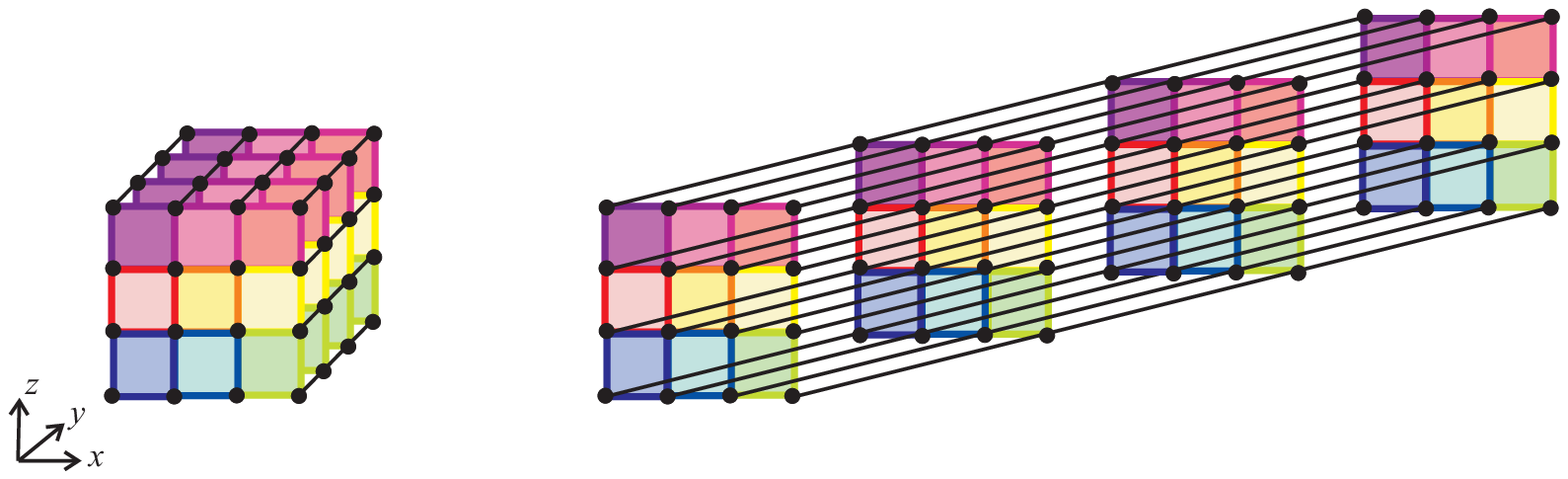}}\\
\hspace{1cm}(a) & \hspace{1.5cm}(b)\\
\end{tabular}
\caption{(a) 3D square lattice. (b) Projection of this 3D square lattice in a plane. Colourful squares correspond to the same squares in figures~\ref{fig:4colourfulsquares+in2Dlattice-colorsof6squares}, \ref{fig:concatenation-embedding-6squares} and \ref{fig:concatenation-embedding-obc}. All edges are decorated even though the decorations are not represented.}
\label{fig:3D-3Dprojected2D-allcolors}
\end{figure}

Next we consider a figure consisting of 4 concatenated faces, each of them with 9 squares (figure~\ref{fig:concatenation-embedding-obc}(a)) with open boundary conditions. We embed this figure in the 2D lattice with crossings as shown in figure~\ref{fig:concatenation-embedding-obc}(b).

It only remains to be shown that such concatenation of faces corresponds to a 3D square lattice projected in a plane. It indeed corresponds to a 3D square lattice if one projects this lattice so that the faces parallel to the plane of the paper do not cross each other, and hence the connections between the faces involve crossings of edges (see figure~\ref{fig:3D-3Dprojected2D-allcolors}). In this manner we obtain a systematic way to identify a 3D square lattice with a non--planar 2D figure.

Thus, we have shown an explicit, constructive
way to write the Hamilton function of the Ising model on a 3D square lattice by writing another Hamilton function on an (enlarged) 2D square lattice with complex parameters, or on a 2D square lattice with crossings with real parameters.

\emph{Polynomial overhead in the system size.} We want to see that the 2D square lattice with crossings is only polynomially enlarged w.r.t the 3D square lattice. To see this, first consider a face of squares of dimensions $x\times z$, with the axes as shown in figure~\ref{fig:3D-3Dprojected2D-allcolors}. Here $x$ ($y,z$) is the number of vertices that the 3D square lattice has in the $x$ ($y,z$) direction (not the number of squares).
Each of the squares has $x-1$ crossings in each of its vertical sides (e.g.  figure~\ref{fig:concatenation-embedding-6squares}(a) shows two faces of squares, each face of dimensions $4 \times 3$, and where each of the squares involves 3 crossings in each of its vertical dimensions). The embedding of such face in a 2D square lattice with crossings (figure~\ref{fig:concatenation-embedding-6squares}(b)) involves $2x+1$ vertices in the horizontal direction and $(x+1)(z-1)$ vertices in the tilted direction going from the lower right corner to the upper left corner (counting the merging at the upper part of the tilted squares). Thus such face requires $(2x+1)(x+1)(z-1)$ vertices in the 2D square lattice with crossings.

The concatenation of faces only adds the additional vertices on which thick black lines go through without having any face at the background (lower right corner and upper left corner in figure~\ref{fig:concatenation-embedding-6squares}(b)). In particular, every concatenation of 2 faces requires $(2x+1)(x+1)$ additional vertices. Therefore the embedding of a 3D square lattice of dimensions $x\times y \times z$ involves $(2x+1) (x+1) [ (z-1) y+ (y-1)] \sim $ vertices in the 2D square lattice with crossings.
Finally, the 2D square lattice with crossings is produced from an enlarged 2D square lattice as summarized in Fig~\ref{fig:concatenation-embedding-obc}(c). We see that the 2D lattice is enlarged by a factor of $4 \times 4 = 16$ w.r.t the 2D lattice with crossings. Hence, we require
\begin{equation}
16 (2x+1) (x+1) [ (z-1) y+ (y-1)] \sim \mc{O}(x^2yz)
\label{eq:overhead3Dto2D}
\end{equation}
vertices to prepare a 3D lattice of size $x\times y \times z$. This proves that the number of qubits of the 2D lattice is polynomially enlarged w.r.t the number of qubits of the 3D square lattice $n=xyz$. In particular, if one considers a 3D square lattice with $x=y=z$, then size of the 2D square lattice grows as $n^{4/3}$. For example, to construct a 3D lattice of size $x=y=z=4$ we need a 2D square lattice with 10800 vertices, i.e.~a lattice of $104\times 104$ qubits. But this is only the number of vertex qubits, to which one has to add the number of edge qubits. Since a 2D lattice with crossings of size $x'\times x'$ with periodic boundary conditions has  $3(x'-1)x'$ edges, and this side has $x'=104$, the required 2D lattice with crossings has in total
$42936$ qubits.

The procedure to project a 3D square lattice on a 2D plane shown in figure~\ref{fig:3D-3Dprojected2D-allcolors} provides a polynomial upper bound in the number of crossings needed to represent a 3D square lattice in a 2D (non--planar) figure. More precisely, for a 3D square lattice of dimensions $x\times y \times z$ the number of crossings $c$ involved in the projection of this figure is $c = x (x-1) (y-1) (z-1) \sim \mathcal{O}(x^2 yz)$.
For example, the 3D square lattice of figure~\ref{fig:3D-3Dprojected2D-allcolors}(a) has dimensions $4\times 4 \times 4$, and its projection involves $c$=108 crossings, as shown in figure~\ref{fig:3D-3Dprojected2D-allcolors}(b). Therefore the number of crossings involved in this projection also grows polynomially with the 3D square lattice and with the same exponent $4/3$.

\section{Completeness results with real parameters} \label{sec:completeness-real}

In this section we restrict the parameters of the complete model to be real, and thereby allow for a physical interpretation. We prove that
\begin{itemize}
\item
The 2D Ising model with magnetic fields is complete with real parameters for all Ising models with magnetic fields on planar graphs. The overhead in the system size is polynomial.
\item
The 3D Ising model without magnetic fields (or the Ising model without magnetic fields on a 2D square lattice with crossings) is complete with real parameters for all Ising models with magnetic fields. The overhead in the system size is polynomial.
\end{itemize}
Regarding completeness results with for $q-$state models, we argue that
\begin{itemize}
\item
A 3D $q-$state edge model (or a 2D $2q-$state model) is complete with real parameters for all $q'-$state edge models with pairwise interactions, with $q'\leq q$. The overhead in the system size is polynomial.
\item
A 2D $q-$state vertex model is complete with real parameters for all $q'-$state edge and vertex models with at most 4$-$body interactions, with $q'\leq q$. The overhead in the system size is polynomial.
\end{itemize}

\subsection{Completeness with real parameters of the 3D Ising model without magnetic fields for Ising models with magnetic fields}
\label{ssec:completeness-real-3DIsing}

\emph{Completeness with real parameters of the 2D Ising model with magnetic fields for Ising models with magnetic fields on planar graphs.}
It has been argued in section~\ref{ex:2Dto3D} that only the merge and deletion rule for the state $|\varphi\ra$ correspond to real parameters, and that there is no combination of them that can create a crossing. Consequently, if we start from the Ising model on a 2D square lattice, which is a planar graph, it can only prepare planar graphs under merge and deletion rule. It follows that the 2D Ising model with magnetic fields is complete with real parameters for Ising models on planar graphs. Moreover, the 2D square lattice needs to be polynomially enlarged w.r.t the graph on which the Ising model is defined, because the final state, $|\varphi_G\ra$ with $G$ planar, is a stabilizer state, and thus it can be prepared efficiently starting from $|\varphi_{2D}\ra$~\cite{He06}.

\emph{Completeness of the 3D Ising model without magnetic fields for Ising models with magnetic fields with real parameters.}
In section~\ref{ssec:psi} we have seen that the state $|\psi\ra$ can incorporate magnetic fields in its description by letting a new vertex $v_0$ interact with all other vertices with a coupling strength equal to the magnetic field (figure~\ref{fig:psi+spike}). In order to prove the main results of this section, we need to introduce the merge and the deletion rule for the state $|\psi\ra$, which will be the only two rules corresponding to real parameters. They coincide with the measurement rules for the state $|\varphi\ra$ (see the beginning of section~\ref{ex:2Dto3D}) but without the vertex qubits and generalized to $q-$level systems (see the end of section~\ref{ssec:psi} for the relation between $|\psi\ra$ and $|\varphi\ra$).

\begin{figure}[htb]
\centering
\psfrag{=}{=}
\psfrag{(a)}{(a)}
\psfrag{(b)}{(b)}
\includegraphics[width=0.3\columnwidth]{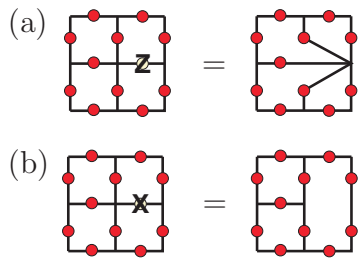}
\caption{Measurement rules for $|\psi\ra$. (a) The merge rule consists of $Z$-measuring the edge qubit. (b) The deletion rule consists of $X$-measuring the edge qubit. Both measurements correspond to real parameters.}
\label{fig:rulesforpsi}
\end{figure}

\emph{Merge rule: $Z-$measurement.}
The merge rule for the state $|\psi_G\ra$ consists of measuring the edge qubit $ab$  in such a way that the resulting state $|\psi_{G'}\ra$ has $a'$ and $b'$ as the same vertex. It coincides with the merge rule for $|\varphi\ra$ in the edge qubit: it consists of projecting the edge qubit on $\la 0|$, i.e.~it is one branch of a $Z$ measurement on the edge qubit (see figure~\ref{fig:rulesforpsi}(a)). Note that this projection corresponds to a coupling strength $h_{0}=\infty$ and $h(i)=-\infty$ for $i\neq 0$, that is, when $s_{a'}-s_{b'}=0$ the coupling is infinitely strong and they effectively become the same particle.

\emph{Deletion rule: $X-$measurement.} It coincides with the deletion of an edge qubit on the state $|\varphi\ra$: it consists of projecting the edge qubit $ab$ onto $\la +_q|$, thus corresponding to one branch of an $X$ measurement
(see figure~\ref{fig:rulesforpsi}(b)). Note that this projection corresponds to a coupling strength $h(j)=0$ for all $j$, that is, the spins $a$ and $b$ have become disconnected.

Now we can proceed with the main results of this section. Here we want to prove that the 3D Ising model \emph{without} magnetic fields (represented by the state $|\psi_{3D}\ra$) is complete with real parameters for Ising models \emph{with} magnetic fields on arbitrary graphs (i.e.~$|\varphi_{G}\ra$ with arbitrary $G$).

\begin{figure}[htb]
\centering
\psfrag{a}{\hspace{-2mm}$|\psi_{3D}\ra$}
\psfrag{b}{\hspace{-2mm}$|\psi_{3D'+h}\ra$}
\psfrag{c}{\hspace{-2mm}$|\varphi_{3D'}\ra$}
\psfrag{d}{\hspace{-3mm}merge$\:$\&}
\psfrag{e}{\hspace{-3mm}\vspace{1mm}deletion}
\psfrag{=}{=}
\includegraphics[width=0.6\columnwidth]{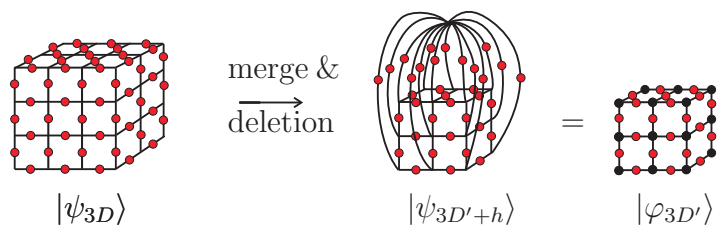}
\caption{The 3D Ising model without magnetic fields ($|\psi_{3D}\ra$) can prepare by means of the merge and deletion rules the 3D Ising model with magnetic fields ($|\psi_{3D'+h}\ra= 2|\varphi_{3D'}\ra$).}
\label{fig:psi3Dtopsi3D+h}
\end{figure}

We first observe that from $|\psi_{3D}\ra$ we can prepare the state  $|\psi_{3D'+h}\ra = 2|\varphi_{3D'}\ra$ by applying only the merge and deletion rules. This state  corresponds to the 3D Ising model with magnetic fields on a smaller lattice (see figure~\ref{fig:psi3Dtopsi3D+h}). To see this, we transform a certain part inside of the 3D square lattice to a 2D square lattice with crossings as shown in section~\ref{ex:2Dto3D}, figure~\ref{fig:concatenation-embedding-obc}. This part will correspond to the smaller 3D square lattice, 3D'. Then we use the remaining part of the 3D square lattice to create one vertex ($v_0$) connected to all vertices of 3D' (black dots in figure~\ref{fig:concatenation-embedding-obc}). Finally the 2D square lattice with crossings is converted to a 3D square lattice again. Note that this procedure only requires merge and deletion rules. Moreover, it shows that there is a polynomial overhead in the number of spins between $|\psi_{3D}\ra$ and $|\psi_{3D'+h}\ra$.
Hence, it only remains to be shown that the 3D Ising model with magnetic fields is complete with real parameters for Ising models with magnetic fields on arbitrary graphs, which is to what we turn next.

Note that, in terms of complexity classes, the 3D Ising model without magnetic fields is known to be as hard as the 2D Ising with magnetic fields (it is an NP-complete problem). However we see that the completeness results \emph{with real parameters} are different for the two models: while the 3D Ising model without magnetic fields is complete for Ising models with magnetic fields on arbitrary graphs, the 2D Ising model with magnetic fields can only prepare Ising models with magnetic fields on planar graphs.

\emph{Completeness of the 3D Ising model with magnetic fields for Ising models with magnetic fields.}
We have seen in section~\ref{ex:2Dto3D} that the state $|\varphi_G\ra$ has to be defined on a non--planar graph in order to prepare $|\varphi_{G'}\ra$ defined on other non--planar graphs if only the merge and deletion rules can be applied. A natural candidate for a non--planar graph is the Ising model with magnetic fields on a 3D square lattice (3D Ising model) which indeed is complete with real parameters for Ising models on arbitrary graphs.

To prove the latter claim, we first observe that a resource is complete if and only if (iff) it can prepare a decorated clique (i.e.~a fully connected graph with all edges being decorated).
The forward implication is trivial, for if a resource can prepare any decorated graph, a particular instance of such decorated graph is a decorated clique. To see the backwards implication, note that, by definition, a decorated clique contains all possible (decorated) edges between the vertices. Thus, in order to obtain a target decorated graph of $n$ vertices one only needs to delete the unnecessary edges of the decorated clique of $n$ vertices by deleting its decoration.

\begin{figure}[htb]
\centering
\psfrag{(a)}{(a)}
\psfrag{(b)}{\vspace{2mm}(b)}
\psfrag{a}{$\tilde{1}$}
\psfrag{b}{$\tilde{2}$}
\psfrag{c}{$\tilde{3}$}
\psfrag{1}{1}
\psfrag{2}{2}
\psfrag{3}{3}
\includegraphics[width=0.6\columnwidth]{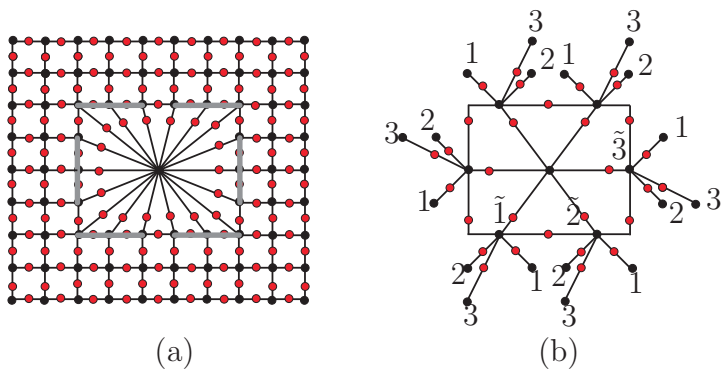}
\caption{Construction of a decorated clique with $n=7$ vertices. Procedure for the lowest layer of the 3D square lattice. (a) First build a hub with $3(n-1)$ neighbors. Delete 2 out of every 3 edges connected to the hub, and merge vertices at the boundary so that only the vertices connected to the hub remain. (b) Use the remaining lattice around the hub to merge vertices to the boundary qubits so that each of them has $n-4$ new neighbors.}
\label{fig:clique-procedure}
\end{figure}

Therefore we need to show that a 3D square lattice can prepare a decorated clique only by merging and deleting qubits.
To construct a decorated clique with $n$ vertex qubits, let us first define the quantities $n_1$, $n_2$ and $n_3$ satisfying $2(n_1+n_2) = n-1$ and $n_3=2(n-4)$. Now we consider a 3D square lattice of size $(3n_1+2n_3)\times (3n_2+2n_3)  \times (n-3)$ (axes shown in figure~\ref{fig:3D-3Dprojected2D-allcolors}) and proceed as follows. We take the lowest layer ($x-y$ plane with $z=1$) and  merge all the vertices inside this layer in order to form a hub connected to all the vertices in the boundary via decorated edges (see figure~\ref{fig:clique-procedure}(a)). We then delete 2 out of every 3 edges that connect the hub to the boundary in a successive way, and merge vertices at the boundary so that only the vertices connected to the hub remain. Notice that the hub has now  $2(n_1+n_2)=n-1$ neighbors in the boundary, and thus this vertex already has all its connections to the other vertices. Now we need to connect the $n-1$ remaining vertices among each other.

\begin{figure}[htb]
\centering
\psfrag{(a)}{(a)}
\psfrag{(b)}{\vspace{2mm}(b)}
\psfrag{a}{$\tilde{1}$}
\psfrag{b}{$\tilde{2}$}
\psfrag{c}{$\tilde{3}$}
\psfrag{1}{1}
\psfrag{2}{2}
\psfrag{3}{3}
\includegraphics[width=0.45\columnwidth]{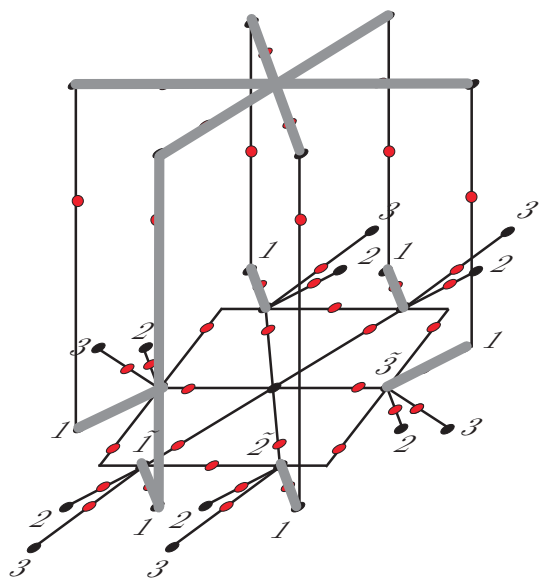}
\caption{All vertices 1 are connected through a hub in the 2nd layer. Gray thick lines indicat edges that have to be merged. In this manner, vertex $\tilde{1}$ is connected to all other vertices. The same procedure is repeated for vertices $2$ through the 3rd layer in order to connect vertex $\tilde{2}$ to all the rest, and so on up to vertices $n-4$.}
\label{fig:clique-procedure-1layer}
\end{figure}

\begin{figure}[htb]
\centering
\includegraphics[width=0.2\columnwidth]{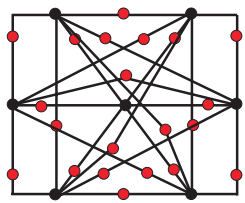}
\caption{A decorated clique of 7 vertices. }
\label{fig:clique}
\end{figure}

To do so, we first merge $n-4$ vertices of the remaining lattice of this layer to each boundary qubit.  Then we delete all other edges in this layer until we obtain figure~\ref{fig:clique-procedure}(b). Now we label boundary qubits as $\tilde{1},\tilde{2},\ldots,\widetilde{n-1}$ and the new neighbors to each boundary qubit as $1, 2, \ldots, n-4$. The idea is that vertex $\tilde{1}$ connects to the other boundary vertices through vertices $1$, and similarly for vertices up to $n-4$. To do so, we connect each vertex 1 to the qubit above it in the 2nd layer ($x-y$ plane, $z=2$). In the 2nd layer we connect them using a hub, all whose connections have to be merged. We also merge each vertex 1 to its boundary qubit, and qubit 1 of the boundary qubit $\tilde{1}$ to its above qubit (see figure~\ref{fig:clique-procedure-1layer}). We repeat the same procedure for qubits 2 through the 3rd layer, and so on up to qubit $n-4$ through the $n-3$ layer. Finally we delete the edges that connect each pair of boundary qubits more than once and we obtain the decorated clique shown in figure~\ref{fig:clique}.

This construction also shows that the 3D square lattice needs to be polynomially enlarged w.r.t the target graph. More precisely, it requires $(3n_1+2n_3)\times (3n_2+2n_3)  \times (n-3)$, with $ 2(n_1+n_2) = n-1$ and $n_3=2(n-4)$, i.e.~$\mc{O}(n^3)$ spins for a target graph with $n$ spins. For example, to construct a clique with $n=7$ vertices as the one shown in figure~\ref{fig:clique}
one requires a 3D lattice of size $10\times 9\times 4 = 270$ vertices.

This proves that the 3D Ising model with magnetic fields is complete with real parameters for Ising models with magnetic fields, and that the overhead in the system size is polynomial.
Due to the remarks at the beginning of this Section, this also proves that the 3D Ising model \emph{without} magnetic fields is complete with real parameters for Ising models with magnetic fields, and the overhead in the system size is also polynomial.

\emph{Completeness of the Ising model without magnetic fields on a 2D square lattice with crossings with real parameters for all Ising models with magneitc fields}.
In the above arguments we have chosen a 3D square lattice as a non--planar graph, but we can as well consider a 2D square lattice with crossings as our starting non--planar graph (such as the one in figure~\ref{fig:summary-example}).
Indeed, a 3D square lattice can be constructed from a 2D square lattice with crossings only by merging and deleting edges as it is shown in Example~\ref{ex:2Dto3D} (there it is shown for the state $|\varphi\ra$ but it also holds for the state $|\psi\ra$ since there also exist merge and deletion rules for $|\psi\ra$ which correspond to real parameters).
Intuitively, they are equivalent because any edge with multiple crossings can be prepared from a 3D square lattice by directly merging this edge through an upper layer, as shown explicitly in section~\ref{ssec:completeness-real-3DIsing}. To prepare the same edge with a 2D square lattice with crossings, we would merge many edges each of which has one crossing, as shown explicitly in Example~\ref{ex:2Dto3D}. Thus, we have effectively ``reduced'' the third dimension (or the upper layer) to a crossing, thereby rendering a non--planar graph which has fewer qubits and can prepare the same non--planar graphs.
This proves that the Ising model without magnetic fields on a 2D square lattice with crossings is complete with real parameters for all Ising models.
Moreover, the 2D square lattice with crossings is still polynomially enlarged w.r.t the target graph, growing with an overhead of $\mc{O}(n^3) \mc{O}(n^{4/3}) \sim \mc{O}(n^{13/3})$.

This result emphasizes the difference between planar and non--planar graphs, instead of 2D and 3D graphs. This difference also arises in terms of complexity, since the Ising model without magnetic fields on a planar graph is in the complexity class P, whereas the same problem on a non--planar graph is NP-hard~\cite{Is00}. 
Other non--planar regular graphs have been considered in the literature, such as the Kuratowskian, which is a 2D lattice with the crossings distributed in a sparser way than ours~\cite{Is00}. Notice that in order to transform these lattices to each other one simply needs to delete some crossings or merge the edges between them.

\subsection{Completeness of 3D $q-$state models for $q-$state models with pairwise interactions}
\label{ssec:completeness-q-real}

Here we show that a $q-$state model on a 3D square lattice is complete with real parameters for all  $q'-$state models, with $q'\leq q$ (both vertex and edge models) with pairwise interactions. The same completeness results hold for a $2q-$state model on a 2D square lattice. In order to prove the results, let us first discuss what measurements applied on the state $|GHZ_q\ra$ (equation~\eref{eq:GHZ-edge}) correspond to real parameters. In particular, we shall see that the merge and deletion rules correspond to real parameters, whereas the creation of a crossing and the preparation of the state $|\chi_i\ra$ (cf. equation~\eref{eq:chi}) require complex parameters.

As before, we express projections in the computational basis and identify the coefficient of $\mathbf{s}^{(i)}$ ($\mathbf{s}^{(i)}=s_a,s_b$ for pairwise interactions) with  $e^{\beta h_i(\mathbf{s}^{(i)})}$ (see~\eref{eq:chi}). We see that only projections with all coefficients real and positive will correspond to real values of the Hamilton function $h_i(\mathbf{s}^{(i)})$.

\begin{figure}[htb]
\centering
\psfrag{(a)}{(a)}
\psfrag{(b)}{(b)}
\psfrag{a}{$\hspace{-1mm}\sum_{i=0}^{q-1}|ii\ra$}
\psfrag{b}{$\hspace{-1mm}|+_q\ra |+_q\ra$}
\psfrag{q}{$q$}
\psfrag{=}{$=$}
\includegraphics[width=0.6\columnwidth]{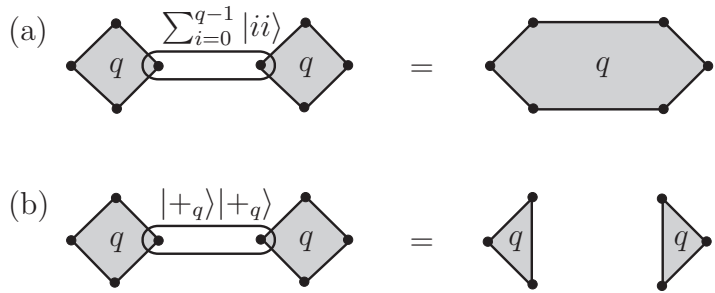}
\caption{(a) Merge rule in the GHZ picture between GHZ states of $q-$level systems. (b) Deletion rule in the GHZ picture.}
\label{fig:merge+delete-GHZ}
\end{figure}

\emph{Merge rule.}
Here the merge rule consists of a certain projection on qubits belonging to two different GHZ states such that they become one single GHZ state. This projection is $| 00 \ra+ | 11\ra  + \ldots +| q-1,q-1\ra $ for GHZ states of $q-$levels (see figure~\ref{fig:merge+delete-GHZ}(a)).
Note that it corresponds to real values of the energy $h_i(\mathbf{s}^{(i)})$. More precisely, it corresponds to assigning $h_i(s_a,s_b)= \infty$ for $s_a=s_b$, and $h_i(s_a,s_b)= -\infty$ otherwise. That is, the coupling between the two is infinitely large when they are in the same state and is infinitely small otherwise, thus forcing the particles to be in the same state and become one single particle.

\emph{Deletion rule.}
This rule consists of deleting the interaction between two particles. This is achieved by the projection of two GHZ states onto $| +_q\ra | +_q\ra$ (see figure~\ref{fig:merge+delete-GHZ}(b)). This corresponds to a vanishing coupling $h_i(\mathbf{s}^{(i)})=0$ for all spin configurations $\mathbf{s}^{(i)}$. The merge and the deletion rule are the only rules in the GHZ picture that correspond to real parameters.

\begin{figure}[htb]
\centering
\psfrag{e}{$l$}
\psfrag{f}{$c$}
\psfrag{g}{$r$}
\psfrag{u}{$u$}
\psfrag{k}{$d$}
\psfrag{i}{$\sum_{i=0}^{q-1}|i\ra_u|i\ra_c$}
\psfrag{b}{$\sum_{i=0}^{q-1}|i\ra_c|i\ra_d$}
\psfrag{d}{$\hspace{-15mm}\sum_{i=0}^{q-1}|i\ra_l |q+i\ra_c$}
\psfrag{c}{$\hspace{-2mm}\sum_{i=0}^{q-1}|q+i\ra_c |i\ra_r$}
\psfrag{q}{$q$}
\psfrag{p}{$\hspace{-1mm}2q$}
\psfrag{=}{$=$}
\includegraphics[width=0.65\columnwidth]{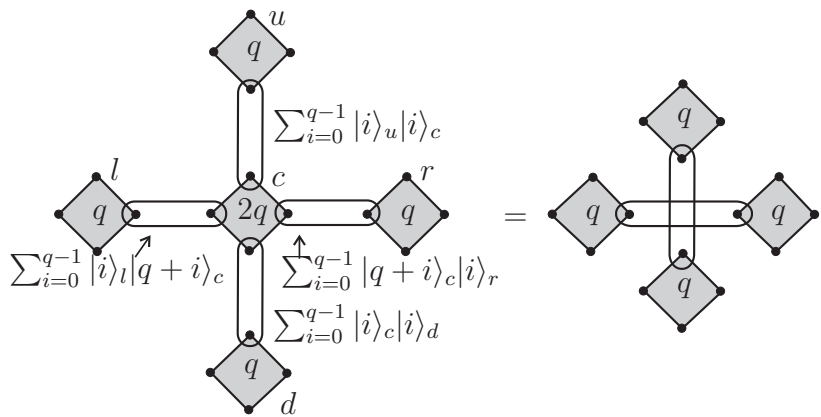}
\caption{Creation of a crossing between $q-$level GHZ states. It requires a non--planar structure if it has to be done with real parameters. The minimal ``non--planar'' structure for a  $q-$level model is a $2q-$level GHZ state. The upper and lowest GHZ states are merged through the lowest levels of this $2q$ GHZ state. The levels of the left particle $l$ are mapped to the upper $q$ levels of $c$, and these are mapped back to the $q$ levels of $r$.}
\label{fig:2q-q}
\end{figure}

\emph{Creation of a crossing.}
Here the creation of a crossed interaction between two pairs of particles is equivalent to generating two merges of interactions that cross each other (see figure~\ref{fig:2q-q} right).
Since the merge and the deletion rules alone cannot generate a crossing, we need to start from a non--planar structure in order to obtain crossings with real parameters. In a system of GHZ states of $q$ levels (see figure~\ref{fig:2q-q}), a GHZ state of $2q$ levels \emph{is} a ``non--planar'' structure because the uppermost $q$ levels can be used to create a crossing in the following way. For the pair of $q-$level GHZ states above and below the central one applies the merge rule going through the $2q-$level GHZ state ($u$, $d$ and $c$ GHZ states of figure~\ref{fig:2q-q}). Now the GHZ states left and right ($l$ and $r$) of the central one have to be merged while crossing the first one.
 To do this, the $q-$levels of $l$ are mapped to the uppermost $q$ levels of $c$, i.e.~the projection reads $\sum_{i=0}^{q-1}|i\ra_l |q+i\ra_c$. Then, these upper levels of $c$ are mapped back to the $q$ levels of $r$, $\sum_{i=0}^{q-1}|q+i\ra_c |i\ra_r$ (see figure~\ref{fig:2q-q}). In this manner, particles $l$ and $r$ are effectively merged. Notice that all projections correspond to real Hamilton functions. That is, we have created a crossing corresponding to real Hamilton functions because of the extra $q$ levels of $c$. Note that this is the minimal ``non--planar'' structure required to create a crossing corresponding to real parameters.

\emph{Preparation of $|\chi_i\ra$}. This procedure does not correspond to a measurement rule, but it is a technique that we used in Secs.~\ref{ssec:completeness-complex-q} and \ref{ssec:2DIsingforvertex} to prove the results. In the preparation of $q-$state models, we prepared the state $|\chi_i\ra= U_i|0\ra^{m_d^{(i)}}$ by constructing a circuit (in the one-way model) that realizes the unitary $U_i$. This circuit generally involves $Y$ measurements as well as tilted measurements not in the range $\cos(\theta/2) |0\ra + \sin(\theta/2) |1\ra$, with $\theta\in [0,\pi ]$, both of which correspond to complex parameters. Therefore, we want to study what gates $U_i$ can be performed with the restriction that the coefficients of all projections are real and positive (i.e.~they are in the above-mentioned range). It seems that one cannot perform any useful gate (such as Hadamard, SWAP, CNOT,...) by applying projections with real and positive coefficients onto $|\varphi_{2D}\ra$. In conclusion, we find that applying (non trivial) gates always corresponds to complex parameters.

Now we can proceed to proving the main results of this section.

\emph{Completeness of 3D $q-$state models with real parameters for all $q-$state models with pairwise interactions}.
In section~\ref{ssec:completeness-complex-q} we mapped models with arbitrary $q$ to an Ising model ($q=2$) applying gates $U_i$ that prepared $|\chi_i\ra$. But we have just seen that these gates require complex parameters, and here we want to obtain completeness results with real parameters. Hence, we take another approach and we consider $q-$state models as candidates to be complete with real parameters for other $q'-$state models, with $q\neq q'$.

In particular, we want to show that a $q-$state model on a 3D square lattice is complete with real parameters for all $q'-$state models on arbitrary graphs, with $q'\leq q$, with pairwise interactions.
We have just seen that the merge and the deletion rule for $q-$state models in the GHZ picture correspond to real parameters.
It follows that all observations regarding these two rules also hold for $q-$state models; in particular, one can transform a 3D square lattice made of $q-$level systems only with the merge and deletion rule into GHZ states of $q-$level systems with an arbitrary interaction pattern, but containing at most pairwise interactions. The reason why we cannot prepare interactions between more particles with real parameters is that their preparation seems to require gates which involve complex parameters. Thus, in order to prepare with real parameters a model with $k-$body interactions, the initial model must contain $k-$body interactions as well.
In summary, if we allow the projections of the GHZs arranged in the 3D square lattice to have its maximum dimension $q^2$, we can prepare with real parameters any other model between $q-$level systems involving at most two$-$body interactions.  Moreover, since these states will be prepared using merge and deletion rules as well, the same efficiency results hold, namely that the 3D $q-$state model is polynomially enlarged w.r.t the target model. This concludes the proof.

\emph{Completeness of a 2D $2q-$state model with real parameters for $q-$state models with pairwise interactions.}
Similarly as we did for Ising models, we realize that we require the 3D structure only because we need to create crossings by merge and deletion rules alone, corresponding to real parameters.
However, we also know that we can use $2q-$level systems in a 2D square lattice to create create a crossing with real parameters between $q-$level systems (figure~\ref{fig:2q-q}). Since we can also use the merge and deletion rules for this 2D $2q-$level system, this proves that  $2q-$level system on a 2D square lattice is complete with real parameters for all $q'-$state models, with $q'\leq q$, with pairwise interactions. Moreover, the overhead in the system size is polynomial, because a polynomially enlarged 2D $2q-$state model can prepare a 3D $q-$state model (as shown in Example~\ref{ex:2Dto3D} for the case $q=2$), and the latter needs to be polynomially enlarged w.r.t the target model.

\subsection{Completeness of 2D $q-$state vertex models for all $q-$state edge and vertex models with at most 4$-$body interactions}
\label{ssec:completeness-vertex-q-real}

Here we show that a $q-$state vertex model on a 2D square lattice (see figure~\ref{fig:peps}) with most general interactions in the vertices is complete with real parameters for $q'-$state vertex and edge models with at most 4$-$body interactions, with $q'\leq q$. Moreover, the 2D $q-$state vertex model needs to be polynomially enlarged w.r.t the target model, be it vertex or edge model.

In order to show that, we need to investigate what measurements can be performed on a vertex model in the PEPS picture that correspond to real parameters. We will see that the situation is different as compared to edge models: not only the merge and deletion rule correspond to real parameters, but also the creation of a crossing can be performed with real parameters. However, the performance of a gate also seems to require complex parameters.

We first note that the coefficients of the projections also have to be identified with the coefficients of~\eref{eq:chi}.

\begin{figure}[htb]
\centering
\psfrag{(a)}{(a)}
\psfrag{(b)}{(b)}
\psfrag{(c)}{(c)}
\psfrag{x}{$\sum_{j=0}^{q-1}| jjjj\ra$}
\psfrag{y}{$| +_q\ra |+_q\ra |+_q\ra |+_q\ra$}
\psfrag{z}{$(\sum_{j=0}^{q-1}|jj|_{ud}\ra )\otimes $}
\psfrag{t}{$(\sum_{k=0}^{q-1}| kk\ra_{lr})$}
\psfrag{q}{$q$}
\psfrag{p}{$\hspace{-1mm}2q$}
\psfrag{=}{$=$}
\includegraphics[width=0.5\columnwidth]{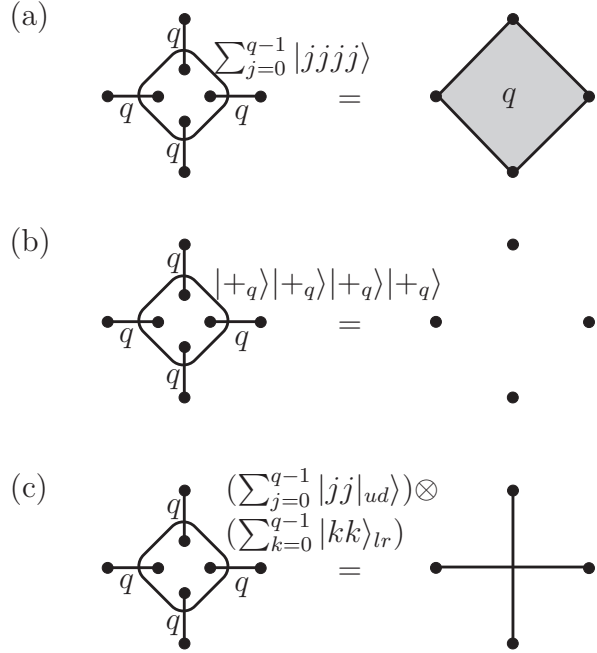}
\caption{Measurement rules in the PEPS picture. (a) Merge rule. (b) Deletion rule. (c) Creation of a crossing. The three rules correspond to real parameters.}
\label{fig:rulesforPEPS}
\end{figure}

\emph{Merge rule}.
By applying the projection $| 0000\ra + | 1111\ra +\ldots +| q-1,q-1,q-1,q-1\ra $ onto four particles belonging to different Bell pairs, we effectively merge them in the sense that they become a ``Bell pair of four particles'', that is to say, a GHZ state of four particles (see figure~\ref{fig:rulesforPEPS}(a)). Note that this measurement rule corresponds to real parameters.

\emph{Deletion rule.} By applying the projection $| +_q\ra |+_q\ra|+_q\ra |+_q\ra$ onto four particles belonging to different Bell pairs, we effectively obtain a deletion rule, in the sense that these four particles become disconnected through this projection, i.e.~there is no interaction (see figure~\ref{fig:rulesforPEPS}(b)). This rule also corresponds to real parameters. The correspondence between each of these rules and the value of the Hamilton function is the same as for the merge and deletion rule for edge models in the GHZ picture (section~\ref{ssec:completeness-q-real}).

\emph{Creation of a crossing}.
To create a crossing, we have to link particles up $u$ and down $d$, and independently, right $r$ and left $l$. Thus the projection $(| 00\ra + | 11\ra +\ldots + | q-1,q-1\ra )_{ud} \otimes ( |00\ra+ | 11\ra \ldots + | q-1,q-1\ra )_{rl}$ creates a crossing (see figure~\ref{fig:rulesforPEPS}(c)). Note that this crossing is created with real parameters and this is possible because the projection itself is between 4 particles.

\emph{Performance of a gate}. In section~\ref{ssec:2DIsingforvertex} we have proven how to map $q-$state vertex models to the 2D Ising model by performing gates $U_i$ which prepare the states $|\chi_i\ra$. However, as we noted in section~\ref{ssec:completeness-q-real}, the performance of gates always seems to correspond to complex parameters.

Now we can proceed to prove the main results of this section.

\emph{Completeness for vertex models.}
To prove that a 2D $q-$state model is complete for vertex models, we note that it can prepare other $q'-$state vertex models (with $q'\leq q$) by merging and deleting edges, and creating crossings, all of which correspond to real parameters. However, if the final model has an interaction between more than 4 particles, this should be prepared using a gate, which involves complex parameters. It follows that in order to prepare (with real parameters) models with $k-$body interactions, the $k-$body interactions must be present in the source model. This is also the reason why we cannot prepare models with $q''> q$: even though the Bell pairs with $q''$ levels can be prepared, an interaction between has to be prepared via a gate, which corresponds to complex parameters. Hence, a 2D $q-$state vertex model with most general interactions in the vertices (i.e.~of dimension $q^4$) can prepare with real parameters any $q'-$state vertex model with $q'\leq q$ with at most 4$-$body interactions.

\emph{Completeness for edge models.}
To prove that a 2D $q-$state model is complete for edge models, we note that it can prepare a GHZ state by merging 4 Bell pairs (see figure~\ref{fig:rulesforPEPS}). Since it can also delete and create crossings with real parameters, it follows that it can prepare any $q'-$state edge model, with $q'\leq q$, with at most 4$-$body interactions.

We remark that in both cases the enlargement of the 2D $q-$state vertex model w.r.t the target model is polynomial, because the models are prepared using the merge, deletion and creation of a crossing rules, and therefore all the previous efficiency results hold.

\section{Relations to complexity theory}
\label{sec:complexity}

In this work we have established reductions from general classes of
classical spin systems to (among others) the 2D and 3D Ising model. More precisely, we have found that the problem of computing the
partition function of a given classical spin model can be seen as a special instance of the problem of computing the partition function of a 2D Ising model with suitably tuned
parameters. In many cases, we have found explicit and efficient reductions
from these models to the 2D Ising model.

The concept of efficiently reducing one (class of) problem(s) to another one
is common practice in the theory of computational complexity. In fact, the
computational complexity of certain standard spin systems, such as e.g. the
Ising and Potts models, is well-studied. In this section we sketch (very
briefly) how the results presented in this paper are related to some of
these established complexity--theoretic results.

The relevant complexity class in the study of partition functions is the
class $\#\mathbf{P}$ (``sharp-P'')~\cite{Pa94}. This class is concerned with
``counting problems''. More specifically, $\#\mathbf{P}$ is defined as
follows: let $f\!\! :\{0, 1\}^N\to\{0, 1\}$ be an arbitrary Boolean function
(representing a decision problem) which can be evaluated in poly$(N)$
time; that is, $f$ represents an arbitrary problem in the class $\mathbf{P}$.
The complexity class $\#\mathbf{P}$ is then concerned with the problem of
computing the integer $\# f$ defined  by $\#f:=|\{x: f(x)=0\}|$. In other
words, problems in $\#\mathbf{P}$ are concerned with \emph{counting} how
many inputs of a given Boolean function yield the output $0$. While such
problems may seem rather innocent at first sight, they are generally very
hard; in fact, there are problems in $\#\mathbf{P}$ which are harder than
all problems in the class $\mathbf{NP}$ (``non-deterministic polynomial
time''), which itself contains numerous problems which are considered to be
intractable. For example, the problem of counting the number of satisfying
assignments to a Boolean 3SAT formula, is a $\#\mathbf{P}$ problem which is
considered intractable (in fact, this problem is known to be
$\#\mathbf{P}$-complete~\cite{Pa94}).

It is well known that the evaluation of the partition function of certain
classes of classical spin systems constitutes a $\#\mathbf{P}$-hard problem
\cite{Ja90}. For example, the computation of the partition function of the 2D Ising model with (inhomogeneous) magnetic fields is known to be
$\#\mathbf{P}$-hard~\cite{Ba82}. This means that every problem in the class
$\#\mathbf{P}$ can be reduced, with polynomial computational effort, to the
computation of the partition function of this model. Similar results have been
found for e.g. the 3D Ising model without magnetic fields, as well as the
Potts models~\cite{Ja90}.

The $\#\mathbf{P}$-hard completeness of the 2D Ising model hence implies
that there exist polynomial reductions from other $\#\mathbf{P}$ problems
(such as the evaluation of partition function of Ising models on arbitrary
graphs, or of the Potts models) to the 2D Ising case. In this sense, the
existence of polynomial reductions from certain spin models to the 2D Ising
case---as we have also reported in the present work---is not a novel
property.  However, although it was indeed known that such reductions exist,
making them explicit is by no means trivial. This is in particular
illuminated by the fact that, for instance, the proof of $\#\mathbf{P}$-hardness of
the 2D Ising partition function is obtained by reducing this problem
problem to an abstract problem from graph theory (the ``MAX CUT'' counting
problem), which is known to be $\#\mathbf{P}$-complete. Such a proof
technique is often used in complexity theory and acts as a kind of
ÒshortcutÓ. Unfortunately, this technique does not seem to give any hint
about the explicit form of the reductions from, say, the 3D Ising model to the
2D case. Thus, it is unclear how useful these complexity results are for
practical purposes.

One of the contributions of the present work is that here these reductions
(e.g.~to the 2D Ising model) are made very explicit: they immediately follow from the
measurement pattern which is needed to produce the respective stabilizer states
from the (decorated) 2D cluster state (and also this measurement pattern can
be determined efficiently). It hence becomes possible to go beyond the
abstract complexity-theoretic statement that the 2D Ising partition function
is $\#\mathbf{P}$-hard, and obtain a result which may hopefully become
 useful in practice.

Another aspect of the present results which does not seem to be covered by
previous complexity-theoretic work, is that the reductions presented here
from general spin systems to, for example, the 2D Ising model, are ``global'' in the
following sense: given a partition function of an Ising model
$Z_{G}(J_{ab}, h_a)$ with pairwise couplings $J_{ab}$ and local fields $h_a$
on an arbitrary graph $G$, we have shown that there exist a 2D lattice
(which is polynomially larger than $G$) and suitable couplings $J_{ij}'$ and
$h_i'$ such that $Z_{G}(J_{ab}, h_a) = Z_{2D}(J_{ij}', h_i)'$ (up to an irrelevant factor). In this
equation, the size of the 2D lattice only depends on $G$ \emph{but not on
the parameters $J_{ab}$ and $h_a$ of the original model}. Moreover, there is
a clear relation between the original parameters $\{J_{ab}, h_a\}$ and
$\{J_{ij}', h_i'\}$. In this sense, we obtain a ``global'' mapping from the
``entire'' spin model on $G$ to the 2D lattice, independent of the couplings
which are considered. As far as we are aware of, this property does not follow
from, say, the $\#\mathbf{P}$-completeness of the 2D Ising model.

\section{Conclusions}
\label{sec:conclusions}

In this paper, we have utilized and extended cross-connections between classical spin models and measurement-based quantum computation. Using tools from quantum information theory, we have shown that there exist mappings between classical spin systems that leave the partition function invariant. With these mappings we have been able to identify {\em complete} models that allow one to express the partition function of {\em any} (discrete) classical spin model as a special instance of the complete model. The mappings are in general between systems with different number of spins, that is, a certain enlargement of the complete model is required so that its partition function can specialize to other partition functions. In all relevant cases, the overhead turns out to be only polynomial.

We believe that one of the main merits of our mappings and completeness results is that they lead to {\em explicit} constructions, which make them also interesting from the viewpoint of complexity theory. We have demonstrated such an explicit construction for the reduction of the 3D Ising model to the 2D Ising model with magnetic fields, where complex parameters are required.

\begin{figure}[htb]
\centering
\psfrag{r}{real}
\psfrag{c}{complex}
\psfrag{2}{2}
\psfrag{3}{3}
\psfrag{D}{$d$}
\psfrag{q}{$q$}
\psfrag{l}{clique}
\psfrag{g}{gates}
\psfrag{k}{$k$}
\psfrag{d}{\hspace{-10mm}complex}
\psfrag{a}{\small{Universality of}}
\psfrag{c}{\small{cluster state}}
\psfrag{b}{\small{Clique}}
\psfrag{A}{(a)\small{ Complex}}
\psfrag{B}{(b) \small{Real}}
\includegraphics[width=0.95\columnwidth]{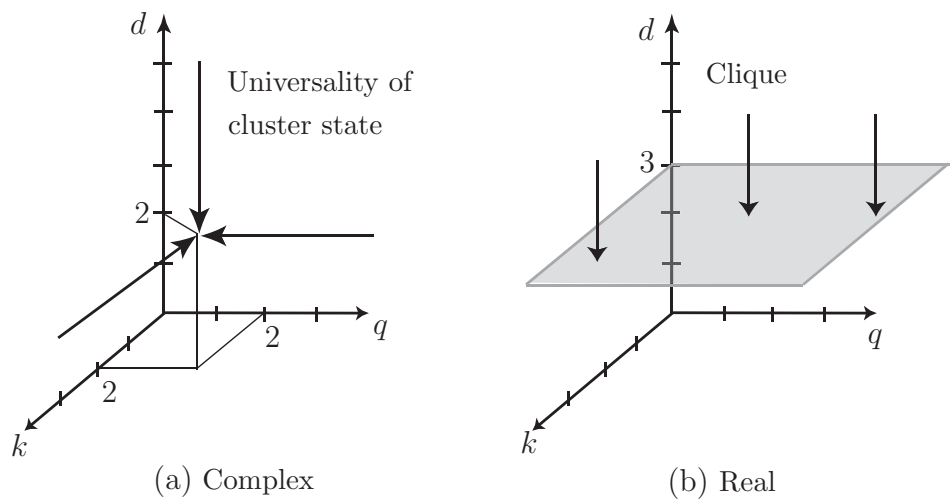}
\caption{The space of all theories: the dimension $d$ versus the number of levels of each particle $q$ versus the number of many body interactions $k$ in the theory. (a) Completeness results with complex parameters. All theories can be reduced to the 2D Ising model (a model with $d=q=k=2$) by invoking the universality of the cluster state.
(b) Completeness results with real parameters. The construction of the clique allows to reduce models with any $d$ to a model with $d=3$, for any $q$ and $k$ fixed. In our constructions, models with higher $q$ or $k$ have to be reached by preparing $|\chi_i\ra$, which involves complex parameters. The ``reductions'' of models to another model with larger $d,q$ or $k$ are trivial.}
\label{fig:spaceofalltheories}
\end{figure}

When allowing for complex parameters, we have found that the 2D Ising model with magnetic fields is complete for all classical spin models, including all $q-$state models with $k-$body interactions, for any $q$ and $k$ (see figure~\ref{fig:spaceofalltheories}(a) for a summary of the completeness results with complex parameters). As long as $k$ and $q$ are bounded, the overhead is polynomial. We have extended our analysis to vertex models, showing the completeness of certain 2D two$-$state vertex models for arbitrary models when allowing for complex parameters.
We have also overcome the limitation of the complex parameters and shown that the 3D Ising model without magnetic field is complete for all Ising models on arbitrary graphs, using only real parameters. This implies that, for these mappings, a physical interpretation is possible, which means that the physical 3D Ising model indeed includes all other Ising models on arbitrary graphs as special instances. In addition, the overhead in these constructions is again only polynomial. Finally, we have shown the completeness of 3D $q-$state vertex models with real parameters for all $q'-$state models with at most 4$-$body interactions and with $q' \leq q$. In summary, in this work we have seen that, with real parameters, we can reduce models of any dimension $d$ to a model with $d=3$, for any \emph{fixed} $q$ and $k$ (see figure~\ref{fig:spaceofalltheories}(b)). However, the reduction to a model with $d=2$, or the reduction of models with high $q$ or $k$ to a model with fixed $q$ or $k$ requires in our constructions the creation of crossings or the preparation of the state $|\chi_i\ra$, respectively, both of which correspond to complex parameters.

Thus, one of the central questions left open in this work is whether a model can be found that is complete for {\em all} classical spin models, including $q-$state models with arbitrary $k-$body interactions, in such a way that all parameters are real. This seems not to be possible for models involving only pairwise interactions, in particular we have not found a way to either obtain larger spin dimensions $q$ or $k-$body interactions with $k>2$ when starting from a two$-$state model with pairwise interactions and restricting ourselves to real parameters. We have, however, recently discovered such a complete model that involves 4$-$body interactions in a 3D lattice~\cite{De08}, where interactions take place between particles on the faces of cubes.

\ack
We thank Liang Jiang and Miguel Angel Martin-Delgado for interesting discussions.
This work was supported by the FWF, the European Union (QICS, SCALA). MVDN acknowledges support by the excellence cluster MAP.

\end{document}